\documentclass[aps,pra,reprint,superscriptaddress,floatfix]{revtex4-2}
\usepackage{graphicx}
\usepackage{amsmath}
\usepackage{amssymb}
\usepackage{amsfonts}
\usepackage{appendix}
\usepackage{mathtools}
\usepackage{leftindex}

\usepackage{newtxtext}
\usepackage{newtxmath}
\usepackage[breaklinks=true,colorlinks,citecolor=blue,linkcolor=blue,urlcolor=blue]{hyperref}
\usepackage[all]{hypcap}
\usepackage{natbib}
\usepackage{multirow}
\usepackage{booktabs}
\usepackage{bbold}
\usepackage{bm}
\usepackage{bbm}

\usepackage{anyfontsize}

\usepackage{indentfirst}
\usepackage{derivative}
\usepackage[dvipsnames]{xcolor}

\usepackage{orcidlink}

\newcommand{\matrixel}[3]{\left\langle{#1}\middle|{#2}\middle|{#3}\right\rangle}
\newcommand{\overlap}[2]{\left\langle{#1}\middle|{#2}\right\rangle}
\newcommand{\trace}[1]{\mathrm{tr}\left({#1}\right)}
\newcommand{\ket}[1]{\left|{#1}\right\rangle}
\newcommand{\bra}[1]{\left\langle{#1}\right|}
\newcommand{\expval}[1]{\left\langle{#1}\right\rangle}

\newcommand{\aop}{\hat{a}}
\newcommand{\adag}{\aop^\dagger}
\newcommand{\adagn}[1]{\aop^{\dagger{#1}}}

\newcommand{\Fop}{\hat{\mathcal{F}}}
\newcommand{\Gop}{\hat{\mathcal{G}}}

\newcommand{\comm}[2]{\left[{#1},{#2}\right]}
\newcommand{\acomm}[2]{\left\{{#1},{#2}\right\}}

\newcommand{\dissip}[1]{\mathcal{D}\left({#1}\right)\left[\rho\right]}

\newcommand{\expen}{\expval{\adag\aop}}

\newcommand{\wss}{W^\mathrm{(ss)}}
\newcommand{\tss}{T_\mathrm{ss}}
\newcommand{\rhoss}{\rho^\mathrm{(ss)}}
\newcommand{\pnss}{P_n^\mathrm{(ss)}}
\newcommand{\enss}{\expval{\adag\aop}^\mathrm{(ss)}}
\newcommand{\woa}{\mathcal{A}}
\newcommand{\wob}{\mathcal{B}}
\newcommand{\woc}{\mathcal{C}}

\newcommand{\rclc}{r_\mathrm{c, lc}}

\newcommand{\moyal}[2]{\left\{\left\{{#1},{#2}\right\}\right\}}

\newcommand{\pleft}{\overset{\leftarrow}{\partial}}
\newcommand{\pright}{\overset{\rightarrow}{\partial}}

\newcommand{\rhoinit}{\rho^\mathrm{(0)}}
\newcommand{\rhoL}{\rho^\mathrm{(L)}}
\newcommand{\rhoR}{\rho^\mathrm{(R)}}
\newcommand{\hsnorm}[2]{\left\langle{#1},{#2}\right\rangle_\mathrm{HS}}

\usepackage[dvipsnames]{xcolor}
\definecolor{RED}{rgb}{1,0,0}
\newenvironment{Revision}{}{}
\newcommand{\revision}[1]{\begin{Revision}{#1}\end{Revision}}

\begin{document}\sloppy
%-------------------------------------------------------------------------------
\title{Transient dynamics of the quantum Stuart-Landau oscillator}
%-------------------------------------------------------------------------------
\author{Hendry Minfui Lim\orcidlink{0000-0002-2957-2438}}
\email[Corresponding author: ]{hendry01@ui.ac.id}
\affiliation{Research Center for Quantum Physics, National Research and Innovation Agency (BRIN), South Tangerang 15314, Indonesia}
\affiliation{Department of Physics, Faculty of Mathematics and Natural Sciences,
Universitas Indonesia, Depok 16424, Indonesia}
%-------------------------------------------------------------------------------
\author{Donny Dwiputra\orcidlink{0000-0002-3645-7892}}
\email[Corresponding author: ]{donny.dwiputra@apctp.org}
\affiliation{Asia Pacific Center for Theoretical Physics, Pohang 37673, Korea}
\affiliation{Research Center for Quantum Physics, National Research and Innovation Agency (BRIN), South Tangerang 15314, Indonesia}
%-------------------------------------------------------------------------------
\author{M~Shoufie Ukhtary\orcidlink{0000-0001-5197-7354}}\email{msho001@brin.go.id}
\affiliation{Research Center for Quantum Physics, National Research and Innovation Agency (BRIN), South Tangerang 15314, Indonesia}
%-------------------------------------------------------------------------------
\author{Ahmad~R.~T.~Nugraha\orcidlink{0000-0002-5108-1467}}\email{ahma080@brin.go.id}
\affiliation{Research Center for Quantum Physics, National Research and Innovation Agency (BRIN), South Tangerang 15314, Indonesia}
\affiliation{Engineering Physics Study Program, School of Electrical Engineering, Telkom University, Bandung 40257, Indonesia}
%-------------------------------------------------------------------------------
\begin{abstract}
%-------------------------------------------------------------------------------
We investigate the transient dynamics of the quantum Stuart-Landau oscillator, a paradigmatic quantum system exhibiting a quantum limit cycle and synchronization.  From the energy dynamics, we determine a condition for the classical regime of transient dynamics and the limit cycle.  Additionally, we formulate a guess function that fits the classical-regime steady-state Wigner function.  The equation of motion for the Wigner function is derived and compared to the Kramers-Moyal equation for stochastic processes.  We then characterize the classical-like behavior as the system evolves from a coherent state, noting the slow decay of neighboring-level coherence.  We also study the evolution of the Wigner negativity as an indicator of nonclassicality, showing its temporary increase for some specific cases.  To quantify the evolution speed, we examined the system's Lindbladian spectra, particularly the Liouvillian gap.  Finally, we record the time it takes to reach the steady state for some Fock, thermal, and coherent states.  The parameter dependence of the steady-state time may differ from the Liouvillian gap, and the limit-cycle attraction is significantly slower for coherent states compared to Fock or thermal states.  For the diagonal states, there are \revision{fast convergence regimes} for which the steady-state time is locally minimized.  This study provides a deeper insight into the transient behavior of self-sustained quantum systems. 
\end{abstract}
% \date{\today}
\maketitle
%-------------------------------------------------------------------------------
\section{Introduction}
%-------------------------------------------------------------------------------
Introducing nonlinearities to a simple harmonic oscillator can make it self-sustaining. A \emph{self-sustained oscillator} is a dynamical system whose phase space exhibits one or more stable trajectories to which a phase point is attracted, corresponding to equilibria between the processes influencing it. It can hold its rhythm against transient perturbations and synchronize via coupling with one another. A self-sustained oscillator is characterized by at least one form of energy gain and energy loss (at least one of which is nonlinear), enabling the equilibrium~\cite{Pikovsky_Rosenblum_Kurths_2001, strogatz2018nonlinear, balanov2008synchronization}. The form of the limit cycle depends on the mathematical description of the system~\cite{HE2005827, ye1986theory}, and multiple limit cycles can exist for more complex systems~\cite{Christopher2024}. Examples of self-sustaining oscillators include wall clocks, electronic systems~\cite{Reddy2005.4524706}, the circadian rhythm~\cite{jewett1999revised, Leloup1999}, chemical reactions~\cite{schnakenberg1979simple, field1974oscillations}, and predator-and-prey systems~\cite{cheng1981uniqueness}. 

The quantum version of self-sustained oscillators has been gaining interest in recent years. One distinct feature of quantum self-sustained oscillators is the spreading of the limit cycle due to quantum noise~\cite{BenArosh2021, Chia2020.PhysRevE.102.042213}, a consequence of the uncertainty principle. Typical mathematical descriptions use the Lindblad master equation~\cite{Rezek2006, Feldmann2004, Rodrigues2007, Walter2014, Chia2020.PhysRevE.102.042213, Benlloch2017.PhysRevLett.119.133601, Dutta2019.PhysRevLett.123.250401, sudler2024.https://doi.org/10.48550/arxiv.2401.03823, marti2023quantum, Sonar2018, Walter2015, Lee2014.PhysRevE.89.022913, Lee2013.PhysRevLett.111.234101, Lorch2016.PhysRevLett.117.073601, Ishibashi2017.PhysRevE.96.052210, Amitai2018.PhysRevE.97.052203, Shen2023, Davis-Tilley_2018}, although different formalisms may be possible~\cite{BOLIVAR2001}. Recent research on quantum self-sustained oscillators is mostly concerned with quantum synchronization, such as quantum synchronization under external driving~\cite{Walter2014, Dutta2019.PhysRevLett.123.250401, Chia2020.PhysRevE.102.042213, marti2023quantum, Benlloch2017.PhysRevLett.119.133601, sudler2024.https://doi.org/10.48550/arxiv.2401.03823, Shen2023}, squeezing~\cite{Sonar2018, marti2023quantum}, and coupling to other oscillators~\cite{Walter2015,Lee2014.PhysRevE.89.022913, Lee2013.PhysRevLett.111.234101, Lorch2016.PhysRevLett.117.073601, Ishibashi2017.PhysRevE.96.052210, Shen2023, Davis-Tilley_2018, Amitai2018.PhysRevE.97.052203}. Intriguing results regarding quantum frequency or phase entrainment, quantum fluctuations~\cite{Lee2013.PhysRevLett.111.234101, Benlloch2017.PhysRevLett.119.133601}, critical response~\cite{Dutta2019.PhysRevLett.123.250401}, information measures~\cite{marti2023quantum}, classicality~\cite{marti2023quantum, Lorch2016.PhysRevLett.117.073601, Sonar2018}, entanglement~\cite{Lee2014.PhysRevE.89.022913}, and amplitude death or oscillation collapse~\cite{Ishibashi2017.PhysRevE.96.052210, Amitai2018.PhysRevE.97.052203, Shen2023} have been reported. Several studies have also proposed quantifying the degree of quantum synchronization~\cite{mari2013, li2017properties, jaseem2020generalized, shen2023qfi.e25081116}, and to speed up synchronization via drive engineering~\cite{Impens2023}.  In addition to quantum synchronization, quantum self-sustained oscillators also play an important role in heat engines~\cite{Rezek2006, Feldmann2004} and single-electron transistors~\cite{Rodrigues2007}. 

Studies have also been done on the dynamics of sole quantum self-sustained oscillators. However, they are mainly concerned with steady-state dynamics, that is, quantum limit cycles, apart from the foundational work of quantizing the equations of motion~\cite{BenArosh2021, Chia2020.PhysRevE.102.042213}. The transient dynamics toward the quantum limit cycle is much less explored. Recently, Ref.~\cite{Dutta2025LimitCycle} studies how a quantum limit cycle may generally emerge, but we believe there is more to uncover. Unlike a classical phase space, richer phenomena like interference, decoherence, and nonclassicality are observable in the quantum phase space. We may anticipate interesting features that are potentially useful for solving related problems, such as those in quantum synchronization. 

In this work, we study the transient dynamics of the quantum Stuart-Landau (SL) oscillator, the paradigm of a quantum self-sustained oscillator exhibiting quantum limit cycle and synchronization. By considering the system's energy evolution, the conditions for the validity of the classical-regime approximation, i.e., the classical-regime eligibility, and the limit cycle's classical regime can be formulated. We also propose a guess function for the classical-regime steady-state Wigner function, which exhibits minimal quantum noise signifying the approach toward the classical limit. 

Moving onto the quantum phase-space representation of the transient dynamics, we derive the equation of motion for the Wigner function. We relate the resulting terms to the features of the Wigner function evolution by comparing them with the Kramers-Moyal (KM) equation for stochastic processes, revealing some deviations due to the quantum nature. We then study the system's evolution starting from a classical-like coherent state and observe how the dynamics vary as the system gains or loses the classical regime eligibility. Additionally, we find that the decay of coherence between the eigenstates of neighboring energy levels is slower than the radial attraction corresponding to the eigenstate occupation probability redistribution. We also consider the negative region of the Wigner function, which signifies nonclassicality. For some initial states, we observe that the Wigner negativity may temporarily increase during the evolution; this effect is attributed to the two-quantum dissipation. 

To characterize the evolution speed, we numerically evaluate the system's Lindbladian spectrum and study its dependence on the system's parameters. We specifically consider the Liouvillian gap, which governs the slowest dynamics toward the limit cycle, noting the false turning point that arises due to bad Hilbert space truncation in the procedure. Lastly, we evolve the system from different initial states (coherent, Fock, thermal) and record the time it takes to reach the steady state. We find that, generally, the evolution does not behave like the Liouvillian gap. For the coherent state, the steady-state time behaves similarly to the Liouvillian gap and does not depend on its initial radial position in phase space. The system takes significantly more time to reach the limit cycle starting as a coherent state than the diagonal states (Fock, thermal), which we attribute to the slow decay of neighboring-level coherence. We observe that the steady-state time is locally minimized over some parameter ranges; these \revision{fast convergence regimes} are observed for the diagonal states, suggesting that they are attributed to the system's energy balance.  

The remainder of this paper is organized as follows. In Section~\ref{section2}, we introduce the mathematical model of the system and discuss some general properties: the connection to the classical SL oscillator, the quantum limit cycle, the disappearance of the phase point due to decoherence, the energy dynamics, and the classical regime. The equation of motion for the system's Wigner function, the features of the evolution of the classical-like coherent state, and the evolution of the Wigner negativity are discussed in Section~\ref{section3}. We consider the Liouvillian gap and characterize the steady-state time for selected initial states in Section~\ref{section4}. Finally, Section~\ref{section5} presents the conclusions and perspectives of our study.

%-------------------------------------------------------------------------------
\section{The System}\label{section2}
%-------------------------------------------------------------------------------

We consider a quantum self-sustained oscillator described by the following Lindblad master equation,
\begin{equation}\label{eq:lme}
\begin{split}
    \rho' &= \mathcal{L}\rho.
\end{split}
\end{equation}
Here, the Liouville superoperator in Lindblad form (or simply Lindbladian) $\mathcal{L}$ is given by
\begin{equation}\label{eq:lindbladian}
\begin{split}
    \mathcal{L}\rho&=-i\comm{\hat{H}_0}{\rho}
    \\
    &\quad +\kappa_1\dissip{\adag}+\gamma_1\dissip{\aop}+\gamma_2\dissip{\aop^2},
\end{split}
\end{equation}
where $\rho'=\mathrm{d}\rho/\mathrm{d}t$, $\hat{H}_0=\adag\aop+1/2$, and $\dissip{\hat{O}}=\hat{O}\rho\hat{O}^\dagger-(1/2)\acomm{\hat{O}^\dagger\hat{O}}{\rho}$. In Eq.~\eqref{eq:lme}, inverse time is measured in units of the oscillator's natural frequency $\omega_0$ (making the simple harmonic period $2\pi$), mass in units of the oscillator's mass $m$, and length in units of $\sqrt{\hbar/m\omega_0}$. This unit system follows Ref.~\cite{BenArosh2021} and makes Eq.~\eqref{eq:lme} dimensionless. Along with $\hbar=1$, the ladder operators $\aop=\left(m\omega_0\hat{x}+i\hat{p}\right)/\sqrt{2\hbar m\omega_0}$ becomes simply $\aop=\left(\hat{x}+i\hat{p}\right)/\sqrt{2}$ in this unit system. There are three sources of nonconservative mechanisms responsible for the self-sustaining property: $\kappa_1$ parameterizes the rate of one-quantum pump, $\gamma_1$ parameterizes the rate of one-quantum dissipation, and $\gamma_2$ parameterizes the rate of two-quantum dissipation. This master equation has been used in the literature, e.g., in Refs.~\cite{BenArosh2021, Dutta2019.PhysRevLett.123.250401, Mok2020}, and is a generalization of another model without the $\gamma_1$ process widely used in the topics of limit cycles and synchronization~\cite{ Walter2014, Chia2020.PhysRevE.102.042213, Benlloch2017.PhysRevLett.119.133601, Sonar2018, Walter2015, Lee2014.PhysRevE.89.022913, Lee2013.PhysRevLett.111.234101, Ishibashi2017.PhysRevE.96.052210, Davis-Tilley_2018}. \revision{In particular, the physical realizations of the nonconservative processes are discussed in Section III A in Ref.~\cite{BenArosh2021}.} Experimental realization of the system in optomechanical setups~\cite{Walter2014, Walter2015, Wachtler2023}, trapped-ion systems~\cite{Lee2013.PhysRevLett.111.234101}, and spin systems~\cite{kato2024quantumspinvander} have been proposed. 

\subsection{The system and the classical Stuart-Landau oscillator}\label{subsection_what_does_the_equation_describe}

The expected phase point dynamics of the system is given by the time dependence of $\expval{\aop}=\trace{\rho\aop}$. Operating Eq.~\eqref{eq:lme} by $\aop$ from either side and using the commutation relation $\comm{\aop}{\adag}=1$, we obtain
\begin{equation}\label{eq:expval_aop_evo}
    \expval{\aop}'=-i\expval{\aop}+\frac{\kappa_1-\gamma_1}{2}\expval{\aop}-\gamma_2\expval{\adag\aop^2}.
\end{equation} 
This equation resembles the equation of motion for the (classical) Stuart-Landau (SL) oscillator~\cite{Pikovsky_Rosenblum_Kurths_2001}:
\begin{equation}\label{eq:expval_aop_evo_semiclassical}
    \alpha'_\mathrm{c}=-i\alpha_\mathrm{c}+\frac{\kappa_1-\gamma_1}{2}\alpha_\mathrm{c}-\gamma_2|\alpha_\mathrm{c}|^2\alpha_\mathrm{c},
\end{equation}
where $\alpha_\mathrm{c}$ is the complex amplitude. This equation follows the same unit system as our quantum system. Using $\alpha_\mathrm{c}\sqrt{2}=r_\mathrm{c}\exp(i\theta_\mathrm{c})$, we separate Eq.~\eqref{eq:expval_aop_evo_semiclassical} into equations for the phase-space radius $r_\mathrm{c}=\sqrt{x_\mathrm{c}^2+p_\mathrm{c}^2}$ and phase $\theta_\mathrm{c}=\tan^{-1}(p_\mathrm{c}/x_\mathrm{c})$:
\begin{subequations}
\begin{align}
    r'_\mathrm{c} &= \frac{(\kappa_1-\gamma_1)r_\mathrm{c}-\gamma_2 r_\mathrm{c}^3}{2}, \label{eq:classical_Stuart_Landau_r}
    \\
    \theta'_\mathrm{c} &= -1.
    \label{eq:classical_Stuart_Landau_theta}
\end{align}
\end{subequations}
The phase point rotates clockwise uniformly at the simple harmonic frequency while being radially attracted toward the circular limit cycle of radius $r_\mathrm{c, lc}=\sqrt{(\kappa_1-\gamma_1)/\gamma_2}$~\footnote{When $\kappa_1<\gamma_1$, the limit cycle radius $r_\mathrm{c, lc}$ becomes imaginary and is thus unphysical. Since the energy pump loses to the linear damping, the two terms in Eq.~\eqref{eq:classical_Stuart_Landau_r} synergize in decreasing $r_\mathrm{c}$ and the oscillation dies out.}, obtained by setting the LHS of Eq.~\eqref{eq:classical_Stuart_Landau_r} to zero and solving for $r_\mathrm{c}$.  If $r_\mathrm{c}<r_\mathrm{c, lc}$, the term $(\kappa_1-\gamma_1)r_\mathrm{c}$ is larger so $r_\mathrm{c}$ increases and vice versa.  This corresponds to the SL oscillator's simple processes to self-sustain its oscillation: (i) a net energy pump when the energy is below the limit-cycle energy, and (ii) a net energy dissipation otherwise. 

%-------------------------------------------------------------------------------
\subsection{The quantum limit cycle}
\label{subsection_toward_the_quantum_limit_cycle}
%-------------------------------------------------------------------------------

The limit cycle of a quantum oscillator is exhibited in the phase space representation of the steady-state density matrix~\cite{BenArosh2021, Chia2020.PhysRevE.102.042213}, analogous to how the limit cycle of a classical oscillator is exhibited in its phase space. In the quantum phase-space representation, we have a quasiprobability distribution that represents the density matrix and, like the density matrix, does not tell us much about the system's observables without being treated under the appropriate mathematical operations~\cite{leonhardt1997measuring, curtright2013concise, schleich2011quantum}. This sets the quantum limit cycle apart from the classical counterpart: the quantum limit cycle is generally \emph{not} just the classical limit cycle spread out into a probability distribution due to the quantum noise. Nevertheless, by observing the Wigner function, we can discern properties of the quantum state that may not be directly evident from the density matrix, e.g., negative regions indicating nonclassicality, entanglement, the expected phase point of a coherent state, and squeezing of the quantum uncertainty.

Ref.~\cite{BenArosh2021} has previously described the transient dynamics in terms of the density matrix $\rho$ and thoroughly studied the steady-state dynamics. Let the $m$th off-diagonal be the density matrix entries $\rho_{n,n+m}$. A given density matrix entry is only coupled to other entries along the same off-diagonal. None of the off-diagonal entries are coupled to the diagonal $(m=0)$ elements and decay to zero, showing that any coherence initially present in the system disappears along the evolution. Meanwhile, the occupation probabilities $P_n=\matrixel{n}{\rho}{n}$ on the density matrix diagonal redistribute toward the steady-state distribution, given by
\begin{equation}\label{eq:pnss}
    \pnss = \frac{\tilde{\kappa}_1^n}{\left(\tilde{\kappa}_1+\tilde{\gamma}_1\right)^{\overline{n}}}\frac{\prescript{}{1}{F_1}(1+n;\tilde{\kappa}_1+\tilde{\gamma}_1+n;\tilde{\kappa}_1)}{\prescript{}{1}{F_1}(1;\tilde{\kappa}_1+\tilde{\gamma}_1;2\tilde{\kappa}_1)},
\end{equation}
where $(x)^{\overline{n}}=x(x+1)(x+2)\dots(x+n-1);\ (x)^{\overline{0}}=1$ is Pochhammer's symbol, $\prescript{}{1}{F_1}(b;c;z)=\sum_{n=0}^\infty[(b)^{\overline{n}}z^n]/[(c)^{\overline{n}} n!]$ is the confluent hypergeometric function~\cite{abramowitz1965handbook}, $\tilde{\kappa}_1=\kappa_1/\gamma_2$, and $\tilde{\gamma}_1=\gamma_1/\gamma_2$. The steady state $\rhoss$, obtained by solving $\mathcal{L}\rhoss=0$, is a statistical mixture of pure states. 

To depict the quantum limit cycle, we use the Wigner function~\cite{leonhardt1997measuring, curtright2013concise, schleich2011quantum}
\begin{equation}\label{eq:wigner_function}
    W(x,p) = \frac{1}{2\pi}\int_{-\infty}^\infty \matrixel{x-\frac{y}{2}}{\rho}{x+\frac{y}{2}}e^{ipy}\ \mathrm{d}y.
\end{equation}
Since $\rhoss$ is diagonal, the corresponding Wigner function $\wss$ is rotationally invariant in the phase space, characterizing the quasiharmonic nature of the limit cycle. Ref.~\cite{BenArosh2021} has previously described some interesting properties of $\wss$. In the regime where $\enss=\trace{\rhoss\adag\aop}\gg 1$, the circular peak of $\wss$ coincides with the classical limit cycle amplitude of the corresponding SL oscillator. As $\enss$ becomes smaller, the peak shrinks in radius, deviating from the classical SL limit cycle. The peak's radius saturates near the zero-point radius $r_\mathrm{zp}=1/\sqrt{2}$ as $\gamma_2\rightarrow\infty$, after which the radius further decreases toward zero as the ratio $\gamma_1/\kappa_1$ gets closer to unity. The Wigner function exhibits \emph{Hopf bifurcation} (i.e., the boundary between the system's limit cycle and damping regime~\cite{strogatz2018nonlinear}) at $\kappa_1=\gamma_1$, similar to the classical SL oscillator. 

%-------------------------------------------------------------------------------
\subsection{Where does the phase point go?}
\label{subsection_where_does_the_phase_point_go}
%-------------------------------------------------------------------------------

The \emph{correspondence principle} dictates that the classical dynamics be restored from the quantum dynamics in the regime where the quantum fuzziness imposed by the uncertainty principle becomes small enough compared to the dynamics of interest that it can be ignored. In this ``classical regime'', any pair of observables becomes compatible, and the uncertainty of any observable becomes negligible. This allows us to write $\comm{\aop}{\adag}\approx 0$ and $\expval{\aop^2}\approx\expval{\aop}^2$, hence $\expval{\adag\aop^2}\approx\expval{\adag}\expval{\aop}^2$, approximating Eq.~\eqref{eq:expval_aop_evo} as Eq.~\eqref{eq:expval_aop_evo_semiclassical} with $\alpha_c \equiv \expval{\aop}$. Starting with a state with $\expval{\aop}\neq 0$, we expect to see a phase point following the classical SL trajectory. However, the quantum equation of motion predicts a steady state $\rhoss$ with a stationary expected phase point $\expval{\aop}^\mathrm{(ss)}=\expval{\adag}^\mathrm{(ss)}=0$, with no assumption about observable compatibility and uncertainty. How come the quantum equation predicts a stationary phase point, while its classical-regime approximation predicts a persistent motion along the limit cycle?

The answer to this question is \emph{quantum decoherence}. Equation~\eqref{eq:lme} takes a simple harmonic oscillator as the system, and the nonconservative processes as the interactions between the system and its environment. As a consequence of ignoring the state of the environment, we progressively lose our knowledge of the system's dynamics. The steady-state density matrix has thus lost track of the expected phase point's motion. As we show in Section~\ref{subsection_the_SL_correspondence_regime}, this is neatly embodied by $\wss$ in the classical regime, which tells us that the expected phase point \emph{may be anywhere} along the classical SL limit cycle, i.e., the circular peak given by $x^2+p^2=(\kappa_1-\gamma_1)/\gamma_2$, with some quantum noise which we may ignore (in the same sense that the Wigner function of a coherent state tells us about its expected phase-point), hence the zero expectation value. In other words, the quantum equation predicts a stationary state that does not ``know'' where the phase point is exactly, but only that it is somewhere on the classical limit cycle.

%-------------------------------------------------------------------------------
\subsection{Energy dynamics and the system's working regimes}
\label{subsection_the_energy_dynamics}
%-------------------------------------------------------------------------------

Repeating the procedure we use to obtain Eq.~\eqref{eq:expval_aop_evo}, but replacing $\aop$ by $\adag\aop\equiv \hat{n}$, we obtain the equation for the evolution of the expected quanta of energy (which we simply call \emph{energy} herein for brevity):
\begin{equation}\label{eq:energy_evo}
\expval{\adag\aop}'=\kappa_1+(\kappa_1-\gamma_1)\expval{\adag\aop}-2\gamma_2\expval{\adagn{2}\aop^2}.
\end{equation}
We can see that the $\gamma_1$ process does not fully oppose the $\kappa_1$ process. The opposition is only for the exponential energy gain, represented by the second term on the right-hand side (RHS) of Eq.~\eqref{eq:energy_evo}. In addition to this increase, the $\kappa_1$ process increases the system's energy linearly with time (the first term on the RHS) unimpeded by the $\gamma_1$ process. This explains why the energy does not simply drop to the vacuum level when $\kappa_1<\gamma_1$~\cite{BenArosh2021}: we still have the linear energy gain from the $\kappa_1$ process competing against the dissipation. On the other hand, the contribution of the $\gamma_2$ process depends on the higher moment $\expval{\adagn{2}\aop^2}$, whose rate of change at any given time depends on its value, the value of $\expval{\adag\aop}$ and the value of the even-higher moment $\expval{\adagn{3}\aop^3}$ at the time, and so on, characterizing the nonlinear nature.

Previous works focused on steady-state dynamics and classified the limit cycle into regimes according to steady-state energy balance, e.g., the ``semiclassical'' $\enss\gg 1$, the ``quantum'' $\enss \approx 1$, and the ``deep quantum'' $\enss \ll 1$ regime.  However, applying the same classification to transient dynamics makes little sense. For example, the ``semiclassical'' limit cycle can be reached by starting in the ``deep quantum'' regime, provided that $\kappa_1\gg\gamma_2$ or $(\kappa_1-\gamma_1)\gg \gamma_2$. We thus need another way to classify the system's overall dynamics, and Eq.~\eqref{eq:energy_evo} is a good candidate. The competing quantities are $\kappa_1$, $(\kappa_1-\gamma_1)$, and $\gamma_2$. Throughout this work, we define the \emph{working regime} of the system using the following ratios: 
\begin{equation}
\begin{split}
\woa = \frac{\kappa_1}{\gamma_2} 
,\qquad
\wob = \frac{\kappa_1-\gamma_1}{\gamma_2}
,\qquad 
\woc = \frac{\woa}{\wob}=\frac{\kappa_1}{\kappa_1-\gamma_1}.
\end{split}
\end{equation}
We have the constraint that $\woc\geq 1$ or $\woa\geq \wob$ above the Hopf bifurcation. Since $\woc$ can be expressed in terms of $\woa$ and $\wob$, either two of $\woa$, $\wob$, $\woc$ suffice to define a \emph{working regime point}. Consequently, we need one of $(\kappa_1,\gamma_1,\gamma_2)$ to serve as the basis for the other two parameters, given the working regime point, to be specified. The working regime points merely make up a \emph{view} of the results when the basis parameter is fixed and the other two parameters are tuned. 

Herein, we use $\kappa_1$ as the basis parameter whenever the working regimes are considered. Increasing $\woa$ means decreasing $\gamma_2=\kappa_1/\woa$; increasing $\woc$ means increasing $\gamma_1=\kappa_1\left(1-1/\woc\right)$; increasing $\wob$ while keeping $\woa$ constant means decreasing $\gamma_1 = \kappa_1\left(1-\wob/\woa\right)$; and so on. While this may seem unnecessarily convoluted, some of the quantities of our interest behave nicely when plotted against the working-regime points. That is, scaling the basis parameter by some constant may leave the values unchanged or scale them by the same constant. Since scaling the basis parameters for a given working regime point means scaling all nonconservative parameters by the same constant, such a property shows the dependence on the \emph{relative} values between the nonconservative parameters rather than their individual values. 

%-------------------------------------------------------------------------------
\subsection{The classical regime eligibility and limit cycle}
\label{subsection_the_SL_correspondence_regime}
%-------------------------------------------------------------------------------

Considering Eq.~\eqref{eq:expval_aop_evo_semiclassical}, the evolution of the energy $E_c=\left|\alpha_c\right|^2$ of the classical SL oscillator is described by
\begin{equation}\label{eq:energy_evo_classical}
\begin{split}
    \left(|\alpha_c|^2\right)' &= \alpha_c'\alpha_c^* + \alpha_c\left(\alpha_c^*\right)' 
    \\
    &= (\kappa_1-\gamma_1)\left|\alpha_c\right|^2-2\gamma_2\left(\left|\alpha_c\right|^2\right)^2.
\end{split}
\end{equation}
The system's energy evolution given in Eq.~\eqref{eq:energy_evo} resembles this equation, except for the linear gain term. This means that the system's energy does not generally evolve resembling its classical counterpart, in the sense that an approximation similar to that in Section~\ref{subsection_where_does_the_phase_point_go} will raise a contradiction. We can interpret the linear gain as a purely quantum contribution, which can be neglected given that
\begin{subequations}
\begin{alignat}{1}
     \woa &\ll 2\expval{\adagn{2}\aop^2}, \label{eq:classical_eligibility_1}
     \\
     C &\ll 
    \expval{\adag\aop}. \label{eq:classical_eligibility_2}
\end{alignat}
\end{subequations}
These constraints define the \emph{classical-regime eligibility} of the system at a given time; the classical approximation does not raise a contradiction if applied in the eligible regime. Since $\expval{\adag\aop}$ and $\expval{\adagn{2}\aop^2}$ at a given time depend on the initial state and the nonconservative parameters, the system may gain or lose eligibility at some point in its evolution toward the limit cycle. 

\begin{figure}[!t]
    \centering
        \includegraphics[trim={0 0.5cm 0 1.5cm},clip, width=\linewidth]{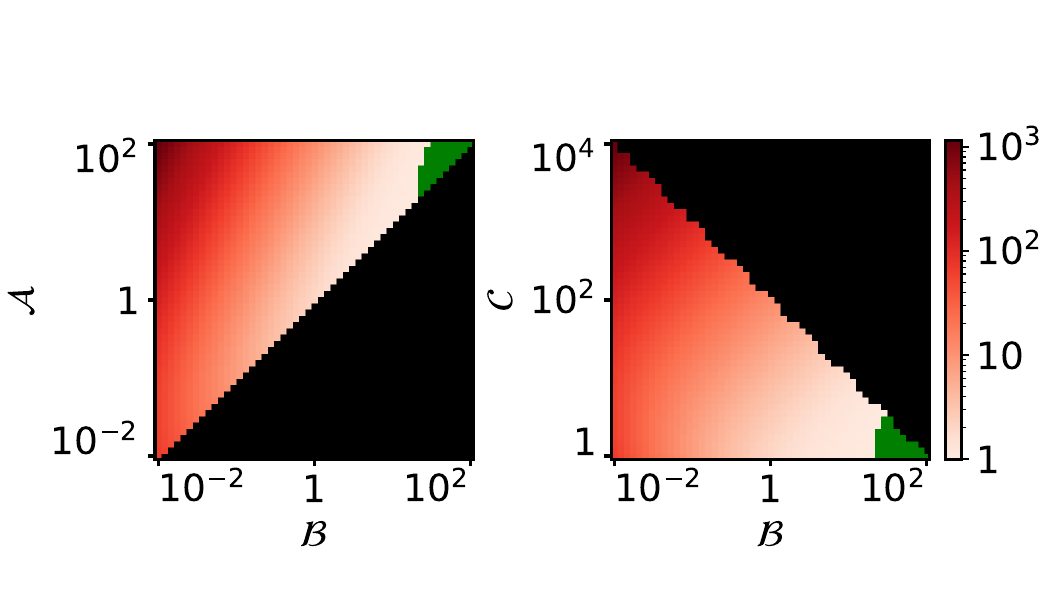}
    \caption{Log-scale plots of the ratio between the steady-state energy $\enss$ and the corresponding SL oscillator's energy $\left|\alpha_\mathrm{c,lc}\right|^2=(\kappa_1-\gamma_1)/2\gamma_2=\wob/2$ in different working regimes. The left and right plots are different views of the same result. The blacked-out regions correspond to the damped regime below the Hopf bifurcation $\woa<\wob$ and to $\woa>100$ for the left and right plots, respectively. The green region marks where the ratio is less than or equal to $1.05$. The ratio never gets smaller than $1$. The basis parameter is $\kappa_1=1$. Scaling $\kappa_1$ by some constant leaves the plots unchanged.}
    \label{fig1}
\end{figure}

Unlike transient dynamics, the steady state is uniquely determined by nonconservative parameters. The parameters are thus sufficient to define the classical regime limit cycle. The energy of the system is determined by the diagonal elements of the density matrix, which do not undergo decoherence, making it a good indicator of the classical regime, i.e., when $\enss\approx|\alpha_\mathrm{c, lc}|^2=(\kappa_1-\gamma_1)/2\gamma_2=\wob/2$. To achieve this, we require the linear energy gain term to be omitted from Eq.~\eqref{eq:energy_evo_classical}, which can be done when 
\begin{equation}\label{eq:classical_regime_limit_cycle}
    \wob \gg 4\woc.
\end{equation}
This defines the \emph{classical regime for the system's limit cycle}. Figure~\ref{fig1} shows the ratio between $\enss$ and $\left|\alpha_\mathrm{c,lc}\right|^2$ in different working regimes. When Eq.~\eqref{eq:classical_regime_limit_cycle} is not satisfied, the linear energy term contributes to the steady-state energy balance, causing $\enss$ to be larger than $\left|\alpha_\mathrm{c, lc}\right|^2$. The smaller $\wob$ is compared to $4\woc$, the more the linear energy term dominates the energy dynamics, and the larger the ratio becomes. As we get close to the bifurcation, i.e., when $\gamma_1\rightarrow\kappa_1$, $\woc$ becomes larger and we need a smaller $\gamma_2$ so that $\wob$ is large enough for the system to enter the classical regime. In this sense, it is harder to enter the classical regime the closer the system is to bifurcation. Interestingly, we find that the ratio for a given $(\woa,\wob)$ is invariant with respect to the choice of basis parameter $\kappa_1$. Since the SL oscillator's energy depends only on $\wob$, this implies that $\enss$ only depends on the working regime variables. 

\begin{figure}[!b]
    \centering
    \includegraphics[width=\linewidth]{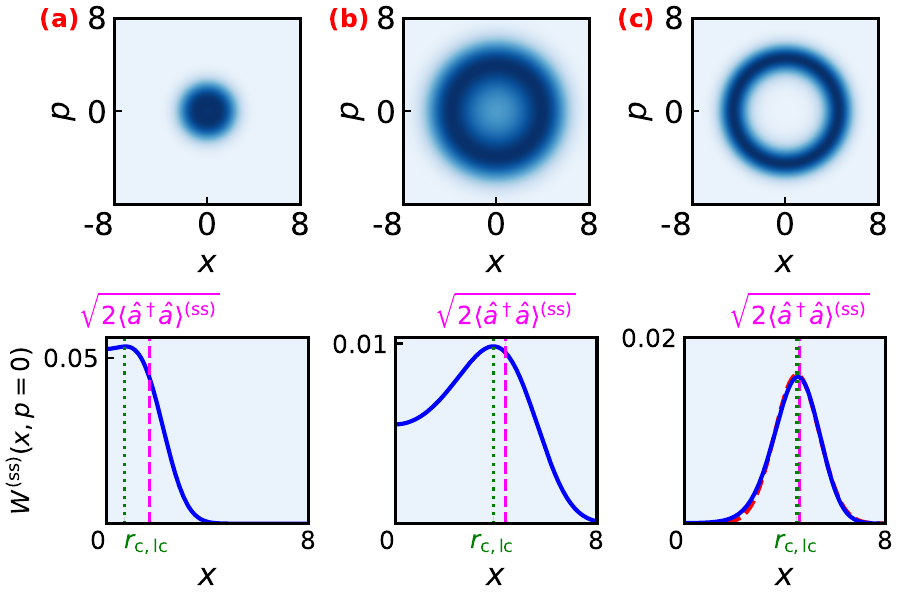}    
    \caption{Steady-state Wigner function $p=0$ cross-section $\wss(x\geq 0,p=0)$ and the corresponding occupation probability distribution $\pnss$ of the first few occupied energy levels, for different parameters $(\kappa_1,\gamma_1,\gamma_2)$: (a) $(1, 0.9, 0.2)$; (b) $(1, 0.9, 0.005)$; (c) $(1, 0.1, 0.045)$. The value of $\enss$ is numerically calculated from $\rhoss$ to be approximately (a) 1.46, (b) 13.08, (c) 10.50. For case (c), the guess function for $\wss$ given by Eq.~\eqref{eq:fit_wss} is additionally plotted as the dashed red line.}
    \label{fig2} 
\end{figure}

Figure~\ref{fig2} shows $\wss$ and its $p=0$ cross-section in different working regimes. As shown in Fig.~\ref{fig2}(a), in the large $\woa$, small $\wob,\woc$ regime we have a clear disparity between the Wigner function peak, $\sqrt{2\enss}$, and $\rclc$. The nonlinear dissipation is strong enough to bring the energy balance down and clump the Wigner function near the origin. By decreasing $\gamma_2$, we get into the regime where the circular peak of $\wss$ coincides with the corresponding SL oscillator's limit cycle. This coincidence occurs even outside the classical regime, as shown in Fig.~\ref{fig2}(b). The parameters do not satisfy Eq.~\eqref{eq:classical_regime_limit_cycle}, so the limit cycle is not in the classical regime. This can also be seen in the Wigner function, where we have a broad distribution toward the origin, not localized enough near the peak to be ignored. The value of $\enss$ also does not agree with $\left|\alpha_\mathrm{c, lc}\right|^2$, as shown by $\sqrt{2\enss}$ that does not coincide with $\rclc=\sqrt{2\left|\alpha_\mathrm{c, lc}\right|^2}$, further compelling us that the peak of $\wss$ in this regime is not to be treated as the limit cycle amplitude.  Consequently, the interpretation that the phase point is somewhere along the Wigner function peak is invalid in this regime.

It is only when the parameters satisfy Eq.~\eqref{eq:classical_regime_limit_cycle} that the limit cycle enters the classical regime, as shown in Fig.~\ref{fig2}(c). Here, $\enss$ approximately agrees with $\left|\alpha_\mathrm{c, lc}\right|^2$ and their corresponding phase-space radii approximately coincide with the peak position of $\wss$, away from the origin. The $\wss$ cross-section approximates a Gaussian distribution centered at $\rclc$. Since the rotationally invariant part of the Wigner function embodies the system's energy (see Appendix~\ref{appsec_A}), the Gaussian shape is necessary for the energy agreement~\footnote{To be more clear, the integral given by Eq.~\eqref{eq:A6} can be transformed into an integral over the phase-space radius times $2\pi$, since the integrand is rotationally invariant. It thus depends only on the cross-section of $\wss$.} and must hence characterize the classical regime limit cycle. As shown by the dashed red line in Fig.~\ref{fig2}(c), we find that $\wss$ is well-fitted by
\begin{equation}\label{eq:fit_wss}
    \wss_\mathrm{guess}(x,p) = \frac{1}{2\pi \rclc} \frac{1}{\sigma\sqrt{2\pi}}\exp\left[-\frac{\left(r-\rclc\right)^2}{2\sigma^2}\right],
\end{equation}
where $r=\sqrt{x^2+p^2}$ and $\sigma^2=\woc/\sqrt{2}$. A mathematical proof for this would be interesting, as we are unable to find one~\footnote{We obtained the expression through the following process. In Fig.~\ref{fig2}(c), we see that the $p=0$ cross-section has a Gaussian shape in the interval $x>0$. Noting the circular symmetry of $\wss$, we can think of the overall Wigner function as revolving this Gaussian curve about an axis perpendicular to the phase-space plane. Since we are considering a (quasi)probability distribution, we consider $\wss(x,p)=\mathcal{N}\left(1/\sigma\sqrt{2\pi}\right)\exp\left[-(r-\mu)^2/2\sigma^2\right]$, where $r=\sqrt{x^2+p^2}$. The normalization condition requires us to add a factor $\mathcal{N}=1/2\pi\rclc$, which is the circumference of the circular peak. Next, the peak of the Gaussian is just $\mu=\rclc$, corresponding to the expected limit cycle radius. Lastly, we need to determine the variance $\sigma^2$ of the distribution. Trying out different sets of parameters, we find that $\sigma^2=\woc/\sqrt{2}$ produces a close approximation. }. Nevertheless, $\wss$ is a surface resulting from the revolution about the origin of a one-variable Gaussian distribution with mean $\rclc$ and variance $\woc/\sqrt{2}$. Since $C\geq 1$, the variance never reaches the limit $1/2$ of the uncertainty principle. While larger $\woc$ increases $\sigma^2$, Eq.~\eqref{eq:classical_regime_limit_cycle} demands that $\wob$ and hence $\rclc$ also increase. Consequently, $\wss$ is always localized enough near its circular peak for us to ignore the quantum fuzziness. It is in this regime (where $\wss$ becomes a bona fide probability distribution) that the interpretation given in Section~\ref{subsection_where_does_the_phase_point_go} makes sense.

%-------------------------------------------------------------------------------
\section{Transient Phase-Space Dynamics}
\label{section3}
%-------------------------------------------------------------------------------

In the previous section, we discussed how we obtain the SL equation in the classical limit, yet our quantum system has a steady state and a stationary expected phase point of zero, and attributed this disagreement to quantum decoherence. How does the system's classical SL behavior vanish as it evolves toward the quantum limit cycle? We also defined the classical regime eligibility, where the classical regime is achievable. How does the system's evolution in the classical regime differ from the other regimes? Furthermore, some quantum states exhibit nonclassical features such as the Wigner function negativity. How do these features change throughout evolution? What are the effects of the nonconservative parameters on these changes? In this section, we characterize how different features of different states change as they evolve toward the quantum limit cycle.

The limit cycle represents the system's dynamics in the limit $t\rightarrow\infty$. Nevertheless, the system can get sufficiently close to the steady state within a finite amount of time. We say that the system is sufficiently close to the steady state at time $t$ when its quantum state distance measure from the steady state drops below a certain threshold. Here we use the trace distance~\cite{Nielsen_Chuang_2010}, for which our criterion reads
\begin{equation}\label{eq:tracedist_to_ss}
    \mathrm{d}_\mathrm{tr}\left(\rho(t),\rhoss\right) = \frac{1}{2} \trace{\left|\rho(t)-\rhoss\right|}\leq \epsilon ,
\end{equation}
where $|\hat{A}|=\sqrt{\hat{A}^\dagger \hat{A}}$. The smallest $t$ for which this inequality is satisfied is what we call \emph{steady-state time} $T_\mathrm{ss}$. We choose the steady-state threshold $\epsilon$ to be small enough that $W$ and $P_n$ are sufficiently close to the steady-state values $T_\mathrm{ss}$, but not too small that the simulations take too long. Choosing $\epsilon=10^{-3}$ achieves this. 

Our results are obtained by evolving the system from appropriate initial states that allow the feature of interest to be present. Some of these states may not be easy to implement for the proposed experimental realizations mentioned in Section~\ref{section2}. However, as initial states, the nonconservative processes need not be present and can be applied after the system is initialized in the given state. The mechanisms used to generate the initial states can thus be whatever, later replaced by our system's nonconservative processes after the initial state is generated. The generation of particular states such as the Fock state~\cite{Simn2020, Wolf2019, Delakouras2023, Weidt2015, Meekhof1996, Tan2014, Huang2019}, the thermal state~\cite{Meekhof1996, An2014, Lee2023}, the coherent state~\cite{Gou1996, Meekhof1996, Zheng1998, Jeon2024, Gerry1997, Alonso2016, li2018}, the squeezed state~\cite{Cirac1993, Zeng1995, Kienzler2017, Meekhof1996, AN2002, Zhu2023, Nunnenkamp2010, Bennett2018, Huang2019, Khosla2017, Youssefi2023, Karg2020}, and the cat state~\cite{Li2023, Hauer2023} has been proposed or even experimentally implemented.

%-------------------------------------------------------------------------------
\subsection{The Wigner function evolution}
\label{subsection_the_wigner_function_evolution}
%-------------------------------------------------------------------------------

The system's Wigner function evolves under $\partial_t W=\hat{\mathcal{D}}_\mathrm{(qSL)}W$, where (see Appendix~\ref{appsec_B} for the derivation details):
\begin{widetext}
\begin{equation}\label{eq:wigner_evo}
\begin{split}
    \hat{\mathcal{D}}_\mathrm{(qSL)} &= -\partial_x p  + \partial_p x 
    \\
    &\qquad -\frac{1}{2}(\kappa_1-\gamma_1)\left( \partial_x x + \partial_p p\right) +\frac{1}{4}(\kappa_1+\gamma_1)\left(\partial_x^2+\partial_p^2\right)
    \\
    &\qquad +\frac{\gamma_2}{2}\left\{ \partial_x \left[\left(x^2+p^2-2\right)x\right]+\partial_p \left[\left(x^2+p^2-2\right)p\right]+\left(\partial_x^2+\partial_p^2\right)\left(x^2+p^2-1\right)+\frac{1}{4}\left(\partial_x^3 x+\partial_x^2\partial_p p + \partial_x\partial_p^2 x + \partial_p^3 p\right) \right\}.
\end{split}
\end{equation}
\end{widetext}
The open system dynamics is stochastic in nature. It thus makes sense that the Wigner function evolves under an equation of motion that resembles the Kramers-Moyal (KM) equation~\cite{gillespie1992markov, risken2012fokker}. The KM equation describes the evolution of a probability distribution $P\left(\{\mathbb{x}\},t\right)$ of a state described by a set $\{\mathbb{x}\}=\{\mathbb{x}_1,\mathbb{x}_2,\dots,\mathbb{x}_N\}$ of Markovian stochastic variables or processes. It is given by
\begin{equation}
    \partial_t P\left(\{\mathbb{x}\},t\right) = \sum_{n=1}^\infty \partial_{\mathbb{x}_{j_1},\dots,\mathbb{x}_{j_n}} D^{(n)}_{\mathbb{x}_{j_1},\dots,\mathbb{x}_{j_n}}\left(\{\mathbb{x}\},t\right)P\left(\{\mathbb{x}\},t\right),
\end{equation}
where $\partial_{\mathbb{x}_{j_1},\dots,\mathbb{x}_{j_n}} = \partial_{\mathbb{x}_{j_1}}\dots \partial_{\mathbb{x}_{j_n}}$ and the KM coefficients are given by
\begin{equation}\label{eq:KM_coefficients}
    D^{(n)}_{\mathbb{x}_{j_1},\dots,\mathbb{x}_{j_n}}\left(\{\mathbb{x}\},t\right) = \frac{(-1)^n}{n!}\lim_{\Delta t\rightarrow 0^+}\frac{1}{\Delta t} \expval{\mathcal{P}^{(n)}_{j_1,\dots,j_n}\left(\Delta t;\{\mathbb{x}\},t\right)}.
\end{equation}
Here the $n$th propagator moment $\mathcal{P}^{n}_{j_1,\dots,j_n}$ of the probability distribution $P$ is given by 
\begin{equation}\label{eq:KM_propagator_moment}
\begin{split}
    \mathcal{P}^{(n)}_{j_1,\dots,j_n}\left(\Delta t;\{\mathbb{x}\},t\right) &= \int_{-\infty}^\infty \mathrm{d}\xi_{j_1}\dots \mathrm{d}\xi_{j_n}
    \\
    &\qquad\qquad \times \xi_{j_1}\dots\xi_{j_n}P\left(\{\mathbb{x}+\xi\},t+\Delta t \right),
\end{split}
\end{equation}
where $\{\mathbb{x}+\xi\} = \{\mathbb{x}_1+\xi_1,\dots,\mathbb{x}_N+\xi_N\}$.

We divide the RHS of Eq.~\eqref{eq:wigner_evo} into three rows. The first row represents the simple harmonic part, which consists of first-order derivatives, resulting in drift terms. Indeed, it is well known that the simple harmonic part is responsible for the rotation of $W$ about the phase-space origin~\cite{curtright2013concise}. 

The second row takes the form of the Fokker-Planck equation, a special case of the KM equation where the RHS goes up to $n=2$~\cite{gillespie1992markov, risken2012fokker}. The first-order derivatives make up the drift term, which shifts $W$ away from (if $\kappa_1>\gamma_1$) or toward (if $\kappa_1<\gamma_1$) the origin, corresponding to the exponential energy gain in Eq.~\eqref{eq:energy_evo}. Meanwhile, the second-order derivatives comprise the constant-rate diffusion terms, which increase the fuzziness of the system and thus embody the quantum decoherence imposed by the one-quantum processes. \revision{This is well-known in the literature (see, e.g., Section 6.3.1 in Ref.~\cite{gardiner2004quantum}).}

So far, the equation of motion satisfies the properties of a KM equation. The third row, however, does not. The Pawula theorem states that the KM equation's RHS ends at either $n=1$, $n=2$, or $n=\infty$~\cite{risken2012fokker}. Furthermore, the coefficients $D^{(n)}_{\mathbb{x}_1,\dots,\mathbb{x}_n}$ is nonnegative for even $n$ and $\mathbb{x}_1=\dots=\mathbb{x}_n$ as evident from Eqs.~\eqref{eq:KM_coefficients} and~\eqref{eq:KM_propagator_moment}. The third row violates these properties. It ends with third-order derivatives, and the second-order coefficient $x^2+p^2-1$ is not nonnegative. This deviation can be justified by noting that $W$ may take on negative values. This means that the propagator moment of $W$ need not be nonnegative for even $n$ and $\mathbb{x}_1 =\dots=\mathbb{x}_n$ and that the Pawula theorem, which assumes that $P\geq 0$, does not apply here. 

We also discuss the continuity of the processes. As pointed out in Chapter 3 of Ref.~\cite{gillespie1992markov}, the KM equation's RHS necessarily truncates at $n=2$ (the Fokker-Planck equation) for a \emph{continuous} Markov process, regardless of whether $P$ is nonnegative. This implies that the Markov process is \emph{jumpy} if the KM equation truncates at $n>2$. In Chapter 4 of Ref.~\cite{gillespie1992markov}, it is shown that for nonnegative $P$ the KM equation never truncates for a jump Markov process. This, however, does not apply in our case, so the previous statement is all we have. We can use these statements to interpret that the energy jumps imposed by the one-quantum processes correspond to small enough jumps in the quantum-phase-space variables that the processes are essentially continuous, resulting in $W$ evolving under a Fokker-Planck equation. Meanwhile, for higher-order energy jumps, the corresponding jumps in the quantum-phase-space variables are not small enough that the resulting contribution to the RHS of Eq.~\eqref{eq:wigner_evo} truncates at $n>2$. In our case of two-quantum dissipation, it truncates at $n=3$. 

\begin{figure*}[!t]
    \centering
    \includegraphics[width=\linewidth]{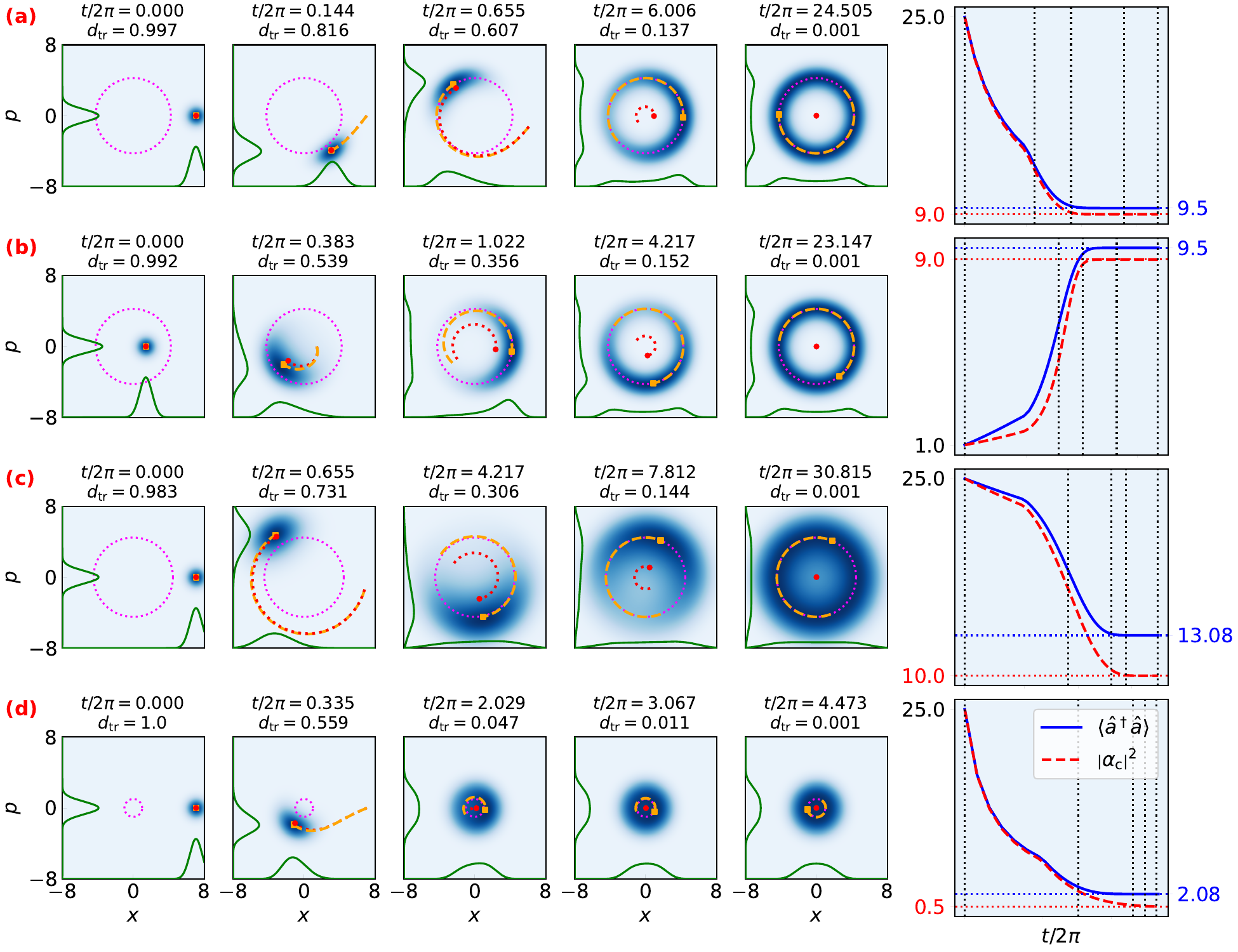}
    \caption{Wigner function $W(x,p)$ representation of the evolution of a coherent state $\ket{\beta}$ under different system parameters $(\kappa_1,\gamma_1,\gamma_2,\beta)$: (a)~$(1, 0.1, 0.05, 5)$; (b)~$(1, 0.1, 0.05, 1)$; (c)~$(1, 0.9, 0.005, 5)$; (d)~$(1, 0.9, 0.1, 5)$; taken at different time points specified on top of each plot over the simple harmonic period of $2\pi$, alongside the trace distance $d_\mathrm{tr}$ toward the steady state given by Eq.~\eqref{eq:tracedist_to_ss}. For each case, the time points are arbitrarily chosen to show: (1) the initial state, (2) some short time into the evolution before the SL phase point reaches its limit cycle, (3) the SL phase point reaching its limit cycle, (4) significant angular spreading of the Wigner function, and (5) when steady state is reached, as indicated by Eq.~\eqref{eq:tracedist_to_ss}. For each plot, the color mapping is normalized against the Wigner function maximum at the given time point. The solid green lines show the Wigner function marginals, where the value 1 is set to be half the plot dimension. The dotted fuchsia circle is the classical SL limit cycle for the same parameters, given by $x^2+p^2=(\kappa_1-\gamma_1)/\gamma_2$. The red dot gives the value of $\expval{a}$ obtained from $\rho(t)$ evolved under Eq.~\eqref{eq:lme}, while the dotted red line is its trail some time into the past. The orange square is the classical SL phase point $\alpha_\mathrm{c}(t)$ obtained by solving Eq.~\eqref{eq:expval_aop_evo_semiclassical}, while the dashed orange line is its trail some time into the past. Along the rightmost column are plots of the expected energy $\expval{\adag\aop}$, and half the radial distance of the classical SL phase point from the origin, which gives the energy of the classical SL oscillator, $E_\mathrm{c}=|\alpha_\mathrm{c}|^2$. The horizontal lines show the stationary values, while the vertical dotted black lines mark the time points of the Wigner function plots along the same row.}
    \label{fig3}
\end{figure*}

We can see the requirement of nonlinearity for the emergence of the limit cycle in Eq.~\eqref{eq:wigner_evo}. For ease of visualization, we consider a vacuum $\rho=\ket{0}\bra{0}$ whose Wigner function is a 2D Gaussian centered at the phase space origin. In this case, the second row only causes the Wigner function to be shifted radially outwards and spread radially uniformly with respect to its peak position at any given time. Without the third row, all we have is a Gaussian spiraling outward while getting spread out. It is the third row that enables non-trivial evolution features resulting in the limit cycle, such as localizing $W$ near the amplitude of the SL limit cycle. Unfortunately, we cannot identify the role of each term due to the unconventional form of the third row.

%-------------------------------------------------------------------------------
\subsection{The classical-like behavior}
\label{subsection_the_classical_like_behavior}
%-------------------------------------------------------------------------------

Under the simple harmonic Hamiltonian $\hat{H}_0=\adag\aop+1/2$, the Wigner function is well-known to rotate about the origin at the angular speed of $1\ \mathrm{rad/s}$. Specifically, for a coherent state $\ket{\beta}=\exp\left(-|\beta|^2/2\right)\sum_n\left(\beta^n/\sqrt{n!}\right)\ket{n}$, the Wigner function peak, which embodies the expected phase point $\expval{\aop}=\matrixel{\beta}{\aop}{\beta}=\beta$, traces out the classical circular trajectory~\cite{curtright2013concise}. Furthermore, its energy is given by $\expval{\adag\aop}=\matrixel{\beta}{\adag\aop}{\beta}=\left|\beta\right|^2$. The coherent state is the closest-to-classical quantum state, which makes it an appropriate choice to study the system's behavior resembling the SL oscillator. We numerically simulate the evolution of the system's Wigner function until it reaches the quantum limit cycle as indicated by Eq.~\eqref{eq:tracedist_to_ss}, starting as a coherent state with different displacements. We also numerically simulate the evolution of the classical SL oscillator's phase point, initialized at the peak position of the coherent state. 

Figure~\ref{fig3} presents the evolution snapshots in different working regimes and the corresponding energy evolution. We generally observe the Wigner function rotating at period $2\pi$ and a radial attraction similar to the SL oscillator. Regardless of the working regime, the radial attraction is similar in that the Wigner function's peak coincides with the SL phase point. This exhibits some form of correspondence between the system and the SL oscillator, even though the peak does not embody the quantity of interest, e.g., the energy. The coincidence can persist until the limit cycle in some regimes (not necessarily the classical regime, as shown in Fig.~\ref{fig2}). In the other regimes, the coincidence is lost as the system evolves.

The quantum decoherence caused by the nonconservative processes takes the form of the angular spreading of the Wigner function resulting from the interplay between the nonconservative processes. Were it absent, we would expect the coherent state to keep its coherence and simply spiral toward the limit cycle amplitude. In reality, this angular spreading causes a gradual loss of information about the phase point, as shown by the expected phase point (red dot and dotted line) deviating from the SL trajectory (orange square and dashed line), and the Wigner function marginals (solid green plots), which shows the position and momentum probability distribution, becoming broader and more symmetric over time. Ultimately, the steady state loses all the information about the phase point, being radially uniform. Nevertheless, the expected phase point agrees with the SL oscillator early in the evolution, regardless of the working regime, thanks to the initial coherence of the coherent state.

By comparing the four cases shown in Fig.~\ref{fig3}, we can see that evolutions in different working regimes are characterized by the angular spreading of the Wigner function. In the classical-eligible regime, the spreading along the radial direction is more confined as it approaches a steady-state distribution localized near the SL limit cycle. This also applies to cases where the system gains classical-regime eligibility somewhere along the evolution. As shown by Fig.~\ref{fig3}(b), the spreading is radially broader early in the evolution but becomes more localized after the system gains eligibility. On the contrary, in Figs.~\ref{fig3}(c) and \ref{fig3}(d), the spreading is more localized early in the evolution but then becomes broader after the system loses eligibility. 

The energy evolution along the rightmost column of Fig.~\ref{fig3} is useful to see where the system loses eligibility and whether the evolution occurs in the classical regime, as indicated by the energy evolution deviating from the SL oscillator. The case of Fig.~\ref{fig3}(a) shows agreement between the system's and the SL oscillator's energy evolution, characterizing that the system's evolution is entirely in the classical regime. This result is interesting: if the system is eligible for all time, then starting as a coherent state, it entirely evolves in the classical regime. The angular spreading happens in such a way that the system's energy keeps agreeing with the SL oscillator. In case (b), the system is initially not eligible, and the energy evolution deviates from the SL oscillator, becoming similar after it gains eligibility and approaching the same limit cycle as case (a). In cases (c) and (d), the system's energy only agrees with the SL oscillator early on, before deviating from it as the system loses eligibility. 

\begin{figure}[!t]
    \centering
    \includegraphics[width=\linewidth]{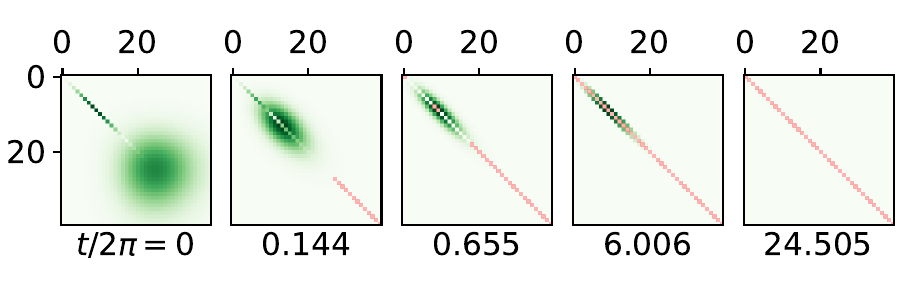}
    \caption{Tile plots showing the difference $\left(\Delta\rho(t)\right)_{mn}$ between the absolute value of the density matrix elements corresponding to case (a) in Fig.~\ref{fig3} and their respective steady-state values, i.e. $\pnss$ as given by Eq.~\eqref{eq:pnss} for the diagonal elements and zero for the off-diagonal elements. The green-gradient color map is normalized against the highest value in the given plot, except when the highest value is lower than $10^{-2}$, in which case it is normalized against $10^{-2}$ instead. The diagonal tiles turn red once the values become smaller than $10^{-3}$.}
    \label{fig4}
\end{figure}

Interestingly, the system's energy reaches the steady-state value well before the system itself reaches the steady state. As shown in Fig.~\ref{fig3}, its Wigner function is yet to be rotationally invariant when $\expen$ gets asymptotically close to $\enss$, indicating leftover coherence when $P_n$ reaches $\pnss$. To see this more clearly, we calculate the difference between the absolute value of each density matrix element and its respective steady-state value: 
\begin{equation}
    \left(\Delta\rho(t)\right)_{mn} = \left|\left|\rho_{mn}(t)\right|-\left|\rhoss_{mn}\right|\right|.
\end{equation}
For case (a) in Fig.~\ref{fig3}, the result is shown in Fig.~\ref{fig4}. The Cauchy-Schwarz inequality warrants that $\left|\rho_{mn}\right|^2\leq \rho_{mm}\rho_{nn}$, i.e., that coherence only exists between two eigenstates with nonzero occupation probability. In our case, the coherence between high-energy eigenstates quickly vanishes. We see a swift shift toward lower-energy eigenstates, following $P_n$ as it quickly shifts due to the dominant nonlinear damping. Afterward, the coherence lingers in the same region as $P_n$ redistributes around $\pnss$. Here, the energy gain is comparable to the dissipation, hence no swift change in $P_n$. As $P_n$ settles down to $\pnss$ in the fourth plot, the leftover coherence is between the eigenstates with nonzero occupation and other eigenstates that are a few energy levels away from them. From numerous simulations with different parameters and $\beta$ values, we find that the coherence between $\ket{n}$ and $\ket{n\pm 1}$ remains large compared to with higher-energy eigenstates, thus taking the longest to vanish. However, the initial magnitude of the $(n,n\pm 1)$ coherence is the largest, so this finding may not apply to an arbitrary quantum state. 

%-------------------------------------------------------------------------------
\subsection{Wigner negativity}
%-------------------------------------------------------------------------------

Departing from the classical-like coherent state, we consider states exhibiting nonclassical features. The Wigner function of some states with nonclassical features may have negative regions. Using the Wigner function, one useful metric for nonclassicality is the negative volume~\cite{Kenfack2004}
\begin{equation}\label{eq:negative_volume}
    \mathcal{V}[W(x,p)] = \frac{1}{2}\left(\int_{-\infty}^\infty\int_{-\infty}^\infty |W(x,p)| \ \mathrm{d}x\ \mathrm{d}p - 1\right).
\end{equation}
Quantum decoherence imposed by the nonconservative processes causes the system to lose its quantumness. This translates to the Wigner function losing its negative regions, resulting in a nonnegative $\wss$ as shown in Fig.~\ref{fig2}. The decoherence imposed by the one-quantum processes is embodied in the terms $\left(\partial_x^2 +\partial_p^2\right)W$ in Eq.~\eqref{eq:wigner_evo}. These terms cause the negative volume to \emph{monotonically decrease}, as we can see by noting that their values near a minimum are positive. Ref.~\cite{Dodonov2005}, for example, has shown this behavior. 

\begin{figure}[!t]
    \centering
    \includegraphics[width=\linewidth]{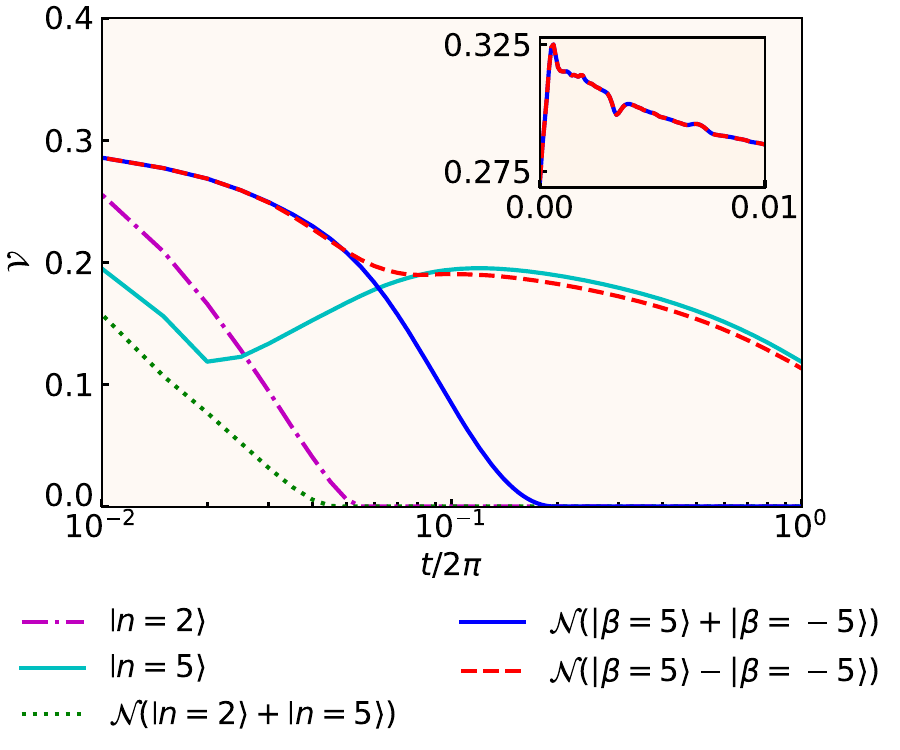}
    \caption{Early evolution of the negative volume $\mathcal{V}$ of the Wigner function given by Eq.~\eqref{eq:negative_volume} of some selected states $\ket{\psi}$, with $(\kappa_1,\gamma_1,\gamma_2)=(0.01,0.009,1)$ and $\mathcal{N}$ some normalization constant. The inset shows the evolution of the cat states' $\mathcal{V}$ for the first hundredth of the simple harmonic period. For the other three states, the evolution of $\mathcal{V}$ over the same time window is simply a monotonous decrease and is not shown. The horizontal axis of the main plot is log-scaled, while that of the inset is linear. A time step of $2\pi/200$ and $2\pi/10000$ is used for the main plot and the inset, respectively. More delicate shapes may be obtained with smaller time steps.}
    \label{fig5}
\end{figure}

On the other hand, the $\gamma_2$ process may decrease $W$ and consequently \emph{increase} $\mathcal{V}$ due to the third-order derivatives in Eq.~\eqref{eq:wigner_evo}. We can see this by making the KM coefficients precede the differential operators~\footnote{As an aside, we can see that the third row is always zero at the phase space origin, meaning that the value of $W$ there never decreases for the system.}. For $\mathcal{V}$ to increase, we need $\gamma_2$ to be large enough compared to $(\kappa_1+\gamma_1)$ so that the third-term decoherence is larger than the second-term decoherence. As such, it is less likely to observe $\mathcal{V}$ increase in the large $\woa$ or $\wob$ regime, such as the classical-eligible regime. Furthermore, we need the third-order derivative values to be negative and have larger magnitudes than the second-order derivative values, limiting which states can exhibit increasing $\mathcal{V}$ to those whose Wigner function's negative regions have quickly decreasing curvatures. Since the terms with second-order derivatives cause $W$ to become less steep with time, it is impossible for $W$ to keep its quickly decreasing curvature. As a result, $\mathcal{V}$ eventually stops being able to increase, and we ultimately have a nonnegative $W$.

Figure~\ref{fig5} shows some examples of $\mathcal{V}$ increasing due to the $\gamma_2$ process in the small $\mathcal{A}$ regime for some initial states showing Wigner negativity. Out of the five states considered, two do not show any increase in $\mathcal{V}$, namely the Fock state $\ket{n=2}$ and the Fock-state superposition $\mathcal{N}\left(\ket{n=2}+\ket{n=5}\right)$, where $\mathcal{N}$ is the normalization constant. While these do not show $\mathcal{V}$ increasing, $\ket{n=5}$ does, suggesting that the Wigner function of $\ket{n=2}$ smoothes that of $\ket{n=5}$. On the other hand, the two cat states show the feature very early in the evolution, suggesting that their sharp interference pattern in the Wigner function is quickly spread out by the dynamics. The different time it takes for $\mathcal{V}$ to vanish depends on the initial states and how the evolution plays out. Here, we solely aim to show that it is possible for $\mathcal{V}$ to increase given the right conditions. 

%-------------------------------------------------------------------------------
\section{Evolution Speed and \revision{fast convergence regimes}}
\label{section4}
%-------------------------------------------------------------------------------

Some interesting features of a self-sustained oscillator are its ability to restore its rhythm after being perturbed and synchronize when coupled. This makes it interesting to study the system's evolution speed. The system becomes more robust to external perturbations by reaching the limit cycle faster. The attraction speed may also provide insight into synchronization enhancement, e.g., in speed and robustness. Our system's dynamics is determined by the nonconservative parameters, as well as the initial state. In this section, we investigate how these affect the time it takes the system to reach the limit cycle. 

%-------------------------------------------------------------------------------
\subsection{The Lindbladian spectra}
%-------------------------------------------------------------------------------

The system's density matrix at any given time can be expressed in terms of the eigenvalues and eigenmatrices of the Lindbladian $\mathcal{L}$ given by Eq.~\eqref{eq:lindbladian}~\cite{Minganti2018, Minganti2019, Haga2021}. Let $\rhoL_j$ and $\rhoR_j$ respectively be the left and right eigenmatrices corresponding to the eigenvalues $\lambda_j$, i.e.
\begin{subequations}
\begin{align}
        \mathcal{L}^\dagger\rhoL_j &= \lambda_j^*\rhoL_j,
        \\
        \mathcal{L}\rhoR_j &= \lambda_j\rhoR_j,
\end{align}
\end{subequations}
where $\mathrm{Re}\left(\lambda_j\right)\leq 0$, and let $\lambda_j$ be sorted in order of ascending magnitudes. If a steady state exists, then $\lambda_0=0$ and $\rhoR_0=\rhoss$ since $\mathcal{L}\rhoss=0$. The density matrix at any given time can be expressed in terms of the eigenvalues and eigenmatrices of $\mathcal{L}$ as
\begin{equation}\label{eq:lindbladian_spectrum_dm_evo}
    \rho(t) = \rhoss + \sum_{j=1}^\infty c_j e^{\lambda_j t}\rhoR_j,
\end{equation}
where $\trace{\rhoR_j}=0$ when $\lambda_j \neq 0$. The spectral decomposition coefficients are given by
\begin{equation}\label{eq:density_matrix_spectral_decomposition_coefficient}
    c_j = \frac{\hsnorm{\rhoL_j}{\rhoinit}}{\hsnorm{\rhoL_j}{\rhoR_j}},
\end{equation}
where $\hsnorm{\hat{A}}{\hat{B}}=\trace{\hat{A}^\dagger\hat{B}}$ is the Hilbert-Schmidt norm and $\rhoinit$ is the initial state. As the system evolves, the real part of $\lambda_j$ causes the summation in Eq.~\eqref{eq:lindbladian_spectrum_dm_evo} to decay toward zero, with the $j=1$ term decaying the slowest. For this reason, $\Delta=\left|\mathrm{Re}\left(\lambda_1\right)\right|$ is called the \emph{Liouvillian gap} or \emph{asymptotic decay rate}. 

The vectorized form of the system's Lindbladian is given by the matrix~\cite{Minganti2018, Minganti2019}
\begin{equation}
\begin{split}
    \bar{\bar{{\mathcal{L}}}} &= -i\left(\adag\aop\otimes \mathbb{1} - \mathbb{1}\otimes\aop\adag\right)
    \\
    &\qquad +\kappa_1\left(\adag\otimes\adag-\frac{1}{2}\aop\adag\otimes\mathbb{1}-\frac{1}{2}\mathbb{1}\otimes\aop\adag\right)
    \\
    &\qquad +\gamma_1\left(\aop\otimes\aop-\frac{1}{2}\adag\aop\otimes\mathbb{1}-\frac{1}{2}\mathbb{1}\otimes\adag\aop\right)
    \\
    &\qquad +\gamma_2\left(\aop^2\otimes\aop^2-\frac{1}{2}\adagn{2}\aop^2\otimes\mathbb{1}-\frac{1}{2}\mathbb{1}\otimes\adagn{2}\aop^2\right),
\end{split}
\end{equation}
whose elements in the energy eigenbasis can be obtained by sandwiching it with the vectorized form $\vec{\rho}_{mn}=\ket{m,n}=\ket{m}\otimes\ket{n}$ of the matrices $\rho_{mn}=\ket{m}\bra{n}$. The matrix elements are given by~\cite{Iachello2024}
\begin{equation}
\begin{split}
    \bar{\bar{{\mathcal{L}}}}_{(a,b);(n,m)} &= \matrixel{a,b}{\bar{\bar{{\mathcal{L}}}}}{m,n}
    \\
    &=\Big[-i(m-n)-\frac{\kappa_1}{2}(m+n+2)-\frac{\gamma_1}{2}(m+n)
    \\
    &\qquad\qquad -\frac{\gamma_2}{2}\left(m(m-1)+n(n-1)\right)\Big]\delta_{a,m}\delta_{b,n}
    \\
    &\qquad +\kappa_1\sqrt{m+1}\sqrt{n+1}\delta_{a,m+1}\delta_{n,m+1}
    \\
    &\qquad + \gamma_1 \sqrt{m}\sqrt{n} \delta_{a,m-1}\delta_{b,n-1}
    \\
    &\qquad + \gamma_2 \sqrt{m(m-1)}\sqrt{n(n-1)}\delta_{a,m-2}\delta_{b,n-2},
\end{split} 
\end{equation}
where $\delta_{m,n}$ is the Kronecker delta. The vectorized form of $\mathcal{L}$ is neither diagonal nor triangular, so the eigenvalues cannot be straightforwardly extracted, and we resort to numerical methods. 

\begin{figure}[!t]
    \centering
    \includegraphics[width=\linewidth]{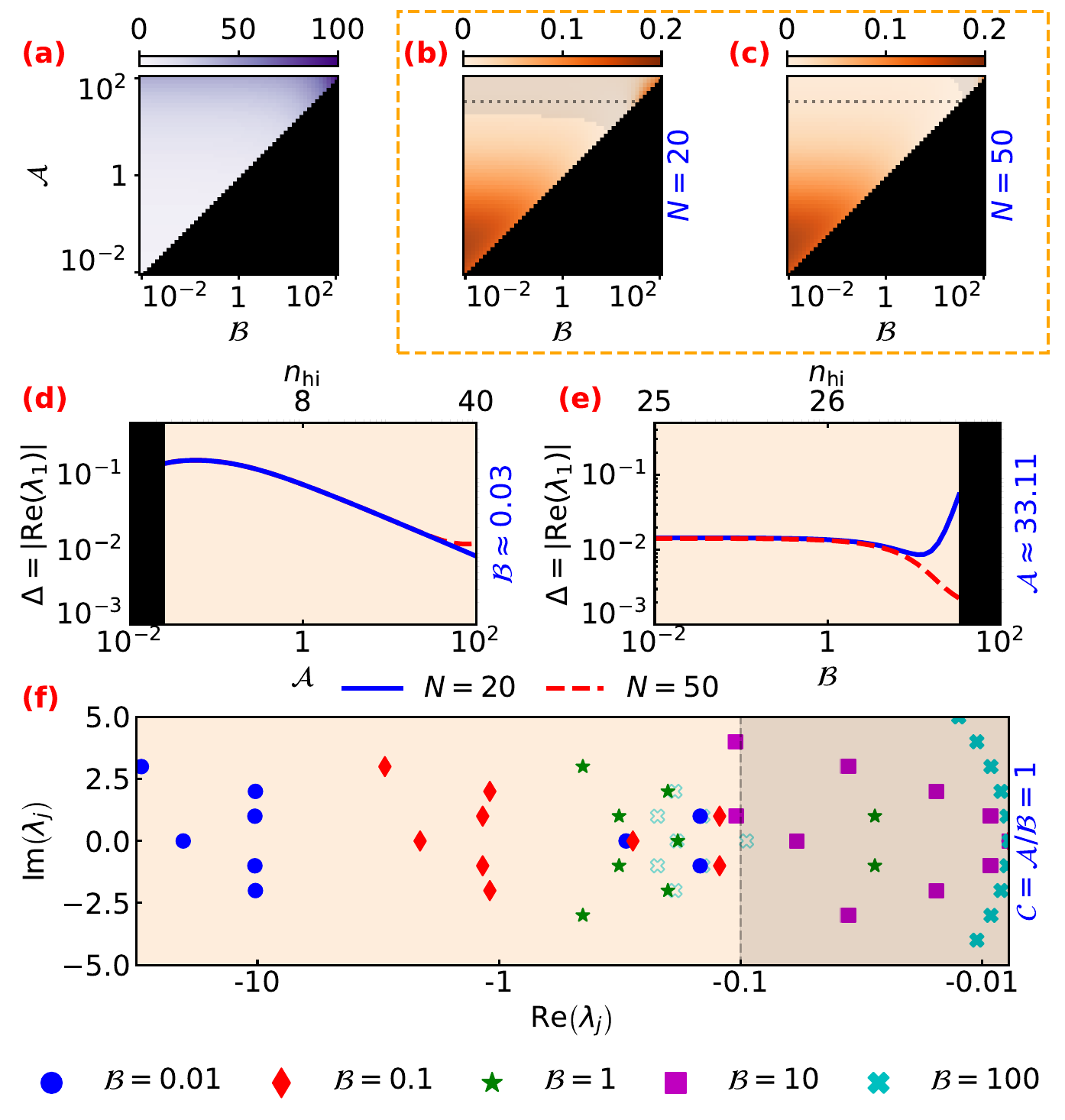}
    \caption{Numerical evaluation of the eigenvalues of the system's Lindbladian in different working regimes. (a): Tile plot of the highest level $n_\mathrm{hi}$ occupied by the steady state $\rhoss$. (b)--(c): Tile plots of the Liouvillian gap $\Delta=\left|\mathrm{Re}\left(\lambda_{1}\right)\right|$ in different working regimes for different truncated Hilbert space dimension $N$. The shaded regions show invalid results: no dynamics from any initial state to the steady state is contained in the first $N$ levels given working regimes points $(\woa=\kappa_1/\gamma_2,\wob=(\kappa_1-\gamma_1)/\gamma_2)$ in the shaded regions. (d)--(e): Plots of $\Delta$ against $\woa$ for $\wob\approx0.03$ and against $\wob$ for $\woa\approx 33.11$. The top $y$-axes show $n_\mathrm{hi}$ for the given $(\woa,\wob)$; the result for a given $N$ is invalid for $x$-axis values for which $n_\mathrm{hi}>N$. (f): Scatter plot of the Lindbladian's eigenvalues $\lambda_j;j=0,1,\dots,9$ in the complex plane for different $\wob$ with $\woc=\woa/\wob=1$. The filled markers correspond to $N=100$, showing a fully valid result since the highest $n_\mathrm{hi}$ is $94$ over the interval. Meanwhile, the empty ones correspond to $N=20$, showing the invalid result most pronounced for $\wob=100$ (at the center of the plot). The $x$-axis is linear between $0$ and $-0.1$ (shaded region), and logarithmic elsewhere. The basis parameter used for the calculations is $\kappa_1=0.1$. Scaling $\kappa_1$ by some constant leaves tile plot (a) and $\mathrm{Im}\left(\lambda_j\right)$ unchanged, and scales $\mathrm{Re}\left(\lambda_j\right)$ by the same constant.}
    \label{fig6}
\end{figure}

While truncating the Hilbert space dimension $N$ means we cannot capture the general dynamics, the results are still valid when the dynamics only involve the first $N$ levels. In Fig.~\ref{fig6}(a), we tile-plot the highest level $n_\mathrm{hi}$ occupied by the system's steady state. Our criterion for this is the lowest level for which $\pnss \leq 10^{-6}$ for all $n_\mathrm{hi} < n\leq N$. The value of $n_\mathrm{hi}$ depends only on $(\woa,\wob)$ and does not depend on the choice of basis parameter. Furthermore, the system's energy monotonously evolves toward $\enss$, as can be seen in, for example, Fig.~\ref{fig3}. These behaviors make $n_\mathrm{hi}$ useful for choosing $N$ to compute the eigenvalues with. We can choose $N$ some levels above $n_\mathrm{hi}$ if the system's dynamics starts at some energy below $\enss$. Otherwise, we may choose $N$ some levels above the highest occupied level of the initial state. Looking at this another way, given the working regime point, say $(\woa,\wob)$, we need at least an $n_\mathrm{hi}$-level approximation for the result to describe the dynamics toward the steady state properly, albeit only for limited choices of initial states. 

Figures~\ref{fig6}(b)--(c) are tile-plots of $\Delta$ in different working regime points and different $N$. In the shaded regions, $n_\mathrm{hi}$ is larger than $N$. As such, the results are invalid there. By increasing $N$, we reveal more valid regions where the values differ from those obtained with a smaller $N$. We can see from the tile plots that a larger $\woa$ leads to a smaller $\Delta$. Other features are less obvious and are easier to see by taking a slice of the tile plots. We take a closer look at the dependence on $\woa$ by taking the $\wob\approx0.03$ slice (6th column from the left) and plotting the values in Fig.~\ref{fig6}(d). For the values plotted, $N=50$ gives the valid result while $N=20$ gives the invalid result. It turns out that in the weakly nonlinear regime (small $\woa$), we can see a small increase in $\Delta$. This increase is then followed by a steady decrease as $\woa$ becomes larger. Meanwhile, for a given $\woa$ we see a plateau-like dependence on $\wob$ which ends at some point as $\wob$ approaches $\woa$, as shown in Fig.~\ref{fig6}(e) for $\woa\approx 33.11$.

As for the validity for a given $N$, contrary to what we described above, the results are pretty similar when $N$ is only a few levels below $n_\mathrm{hi}$ where we find the relative error to be as small as 1.7\%. Nevertheless, as $n_\mathrm{hi}$ becomes significantly larger than $N$, we see a significant deviation of the invalid values. Had we ignored the $n_\mathrm{hi}$ condition, we would have erroneously concluded that the Liouvillian gap can increase as $\gamma_1$ becomes sufficiently smaller than $\kappa_1$, forming a turning point that signifies a phase transition~\cite{Minganti2018}. By increasing $N$, the turning point of $\Delta$ decreases and its position shifts toward larger $\wob$. As such, the true dynamics obtained as $N\rightarrow\infty$ is that there is no such turning point.

Shifting toward a more general viewpoint, in Fig.~\ref{fig6}(f), we scatter-plot the distribution of $\lambda_j;j=0,1,\dots,9$ in the complex plane for different $\wob$ with $\woc=1$. This shows the system's Liouvillian spectra close to Hopf bifurcation (when $\kappa_1=\gamma_1$), whose general features have been previously discussed by Ref.~\cite{Dutta2025LimitCycle}. The opaque and faded markers correspond to $N=100$ and $N=20$, respectively. The highest $n_\mathrm{hi}$ is $94$ in the plotted interval, so the result for $N=100$ is fully valid. Similar to Fig.~\ref{fig6}(e), a large deviation between the valid and invalid results can be seen when $\wob$ becomes large compared to $\woa$. The Liouvillian gap $\Delta$ corresponds to a complex conjugate pair of eigenvalues: $\lambda_1$ and $\lambda_2$. Eigenvalues with larger real-part magnitudes are also multiply real-part degenerate. As $\wob$ (and $\woa$) increases, the eigenvalues decrease in real-part magnitudes at different rates, so real-part degeneracies higher than 2 are progressively lost. At $\wob=100$ and $\woc=1$, the closest-to-zero eigenvalues become doubly degenerate. They lie close to the imaginary axis of the plot, indicating that the system is close to the classical limit, for which the Liouvillian eigenvalues lie on the imaginary axis~\cite{Dutta2025LimitCycle}. Unlike $n_\mathrm{hi}$ and the ratio $\enss/\left|\alpha_\mathrm{c, lc}\right|^2$ shown in Fig.~\ref{fig1}, the real parts of the eigenvalues are scaled along the basis parameter. Meanwhile, the imaginary parts are left unchanged, which is consistent with the fact that the nonconservative processes do not affect the oscillation (i.e., the rotation of the system's Wigner function). This can be seen, for example, in Fig.~\ref{fig3}, which shows the Wigner function rotating in the phase plane at the simple harmonic frequency.

%-------------------------------------------------------------------------------
\subsection{Steady-state times for selected initial states}
\label{subsection_steady_state_time_for_selected_initial_states}
%-------------------------------------------------------------------------------

Despite the nonlinear nature of the system, the change in $\Delta$ over different working regimes is intriguingly simple. Subsequent values of $\left|\mathrm{Re}\left(\lambda_j\right)\right|$ are packed close enough that the spectral decomposition coefficients $c_j$, which additionally depends on the initial state, as given in Eq.~\eqref{eq:density_matrix_spectral_decomposition_coefficient}, may significantly affect the system's transient dynamics. It is thus interesting to study how fast different initial states $\rhoinit$ can actually reach the steady state $\rhoss$, and how different working regimes influence how fast $\rhoss$ is reached for the given $\rhoinit$. 

One measure for the transient lifetime is the relaxation time $\tau$, which measures the time at which the difference between some observable and its steady-state value reaches $1/e$ of the value at the beginning of the evolution~\cite{Haga2021}. We are, however, interested in the actual time it takes for the system to reach $\rhoss$. Adding to this the nonlinear nature of the system, we instead use the steady-state time $\tss$, the criterion for which is given by Eq.~\eqref{eq:tracedist_to_ss}.

Obviously, we cannot study the evolution of all density matrices. Here we consider three kinds of initial states. The first is the Fock state 
\begin{equation}
\rho_\mathrm{Fock}=\ket{n}\bra{n},
\end{equation}
which we choose for its simplicity as an eigenstate of the harmonic oscillator that exhibits Wigner negativity. Next, we consider the thermal state 
\begin{equation}
\rho_\mathrm{th}=\sum_k\frac{\tilde{n}_\mathrm{th}^k}{\left(1+\tilde{n}_\mathrm{th}\right)^{k+1}}\ket{k}\bra{k},
\end{equation}
to study the system's transient limit cycle attraction starting in a thermal equilibrium with an average energy of $\expval{\adag\aop}=\tilde{n}_\mathrm{th}$. Lastly, we study the coherent state, whose density matrix is given by
\begin{equation}
\rho_\mathrm{coh}=\ket{\beta}\bra{\beta} = e^{-|\beta|^2}\sum_{mn}\frac{\beta^m\beta^{*n}}{\sqrt{m!n!}}\ket{m}\bra{n},    
\end{equation}
where $\expval{\adag\aop}=|\beta|^2$, which possesses quantum coherence. To compare different initial states, we start with the same energy $\expval{\adag\aop}^{(0)}$ so the early dynamics is only influenced by the $\gamma_2$ process, as evident from Eq.~\eqref{eq:energy_evo}. This is as far as we can get to putting them on equal footing. We choose $\expval{\adag\aop}^{(0)}=1,3,10$ for each initial state, which we deem sufficient to show the change in the behavior of $\tss$.

\begin{figure}[!t]
    \centering
    \includegraphics[width=\linewidth]{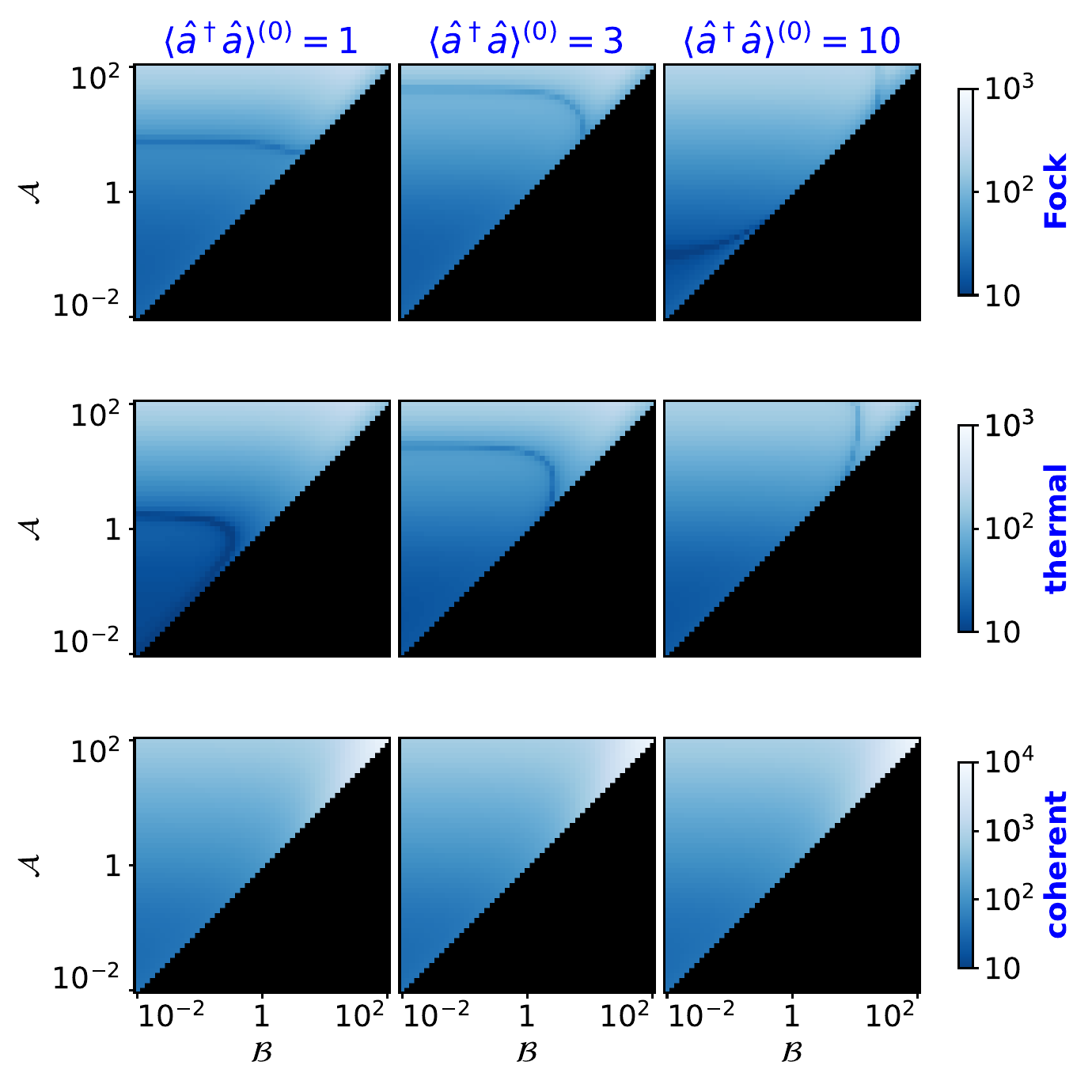}
    \caption{The steady-state time $T_\mathrm{ss}$ for different working regime points for selected initial states with an increasing initial energy $\expval{\adag\aop}^{(0)}$. The basis parameter used is $\kappa_1=0.1$. Scaling $\kappa_1$ by some constant scales the values nonuniformly.}
    \label{fig7}
\end{figure}

\begin{figure}[!t]
    \centering
    \includegraphics[width=\linewidth]{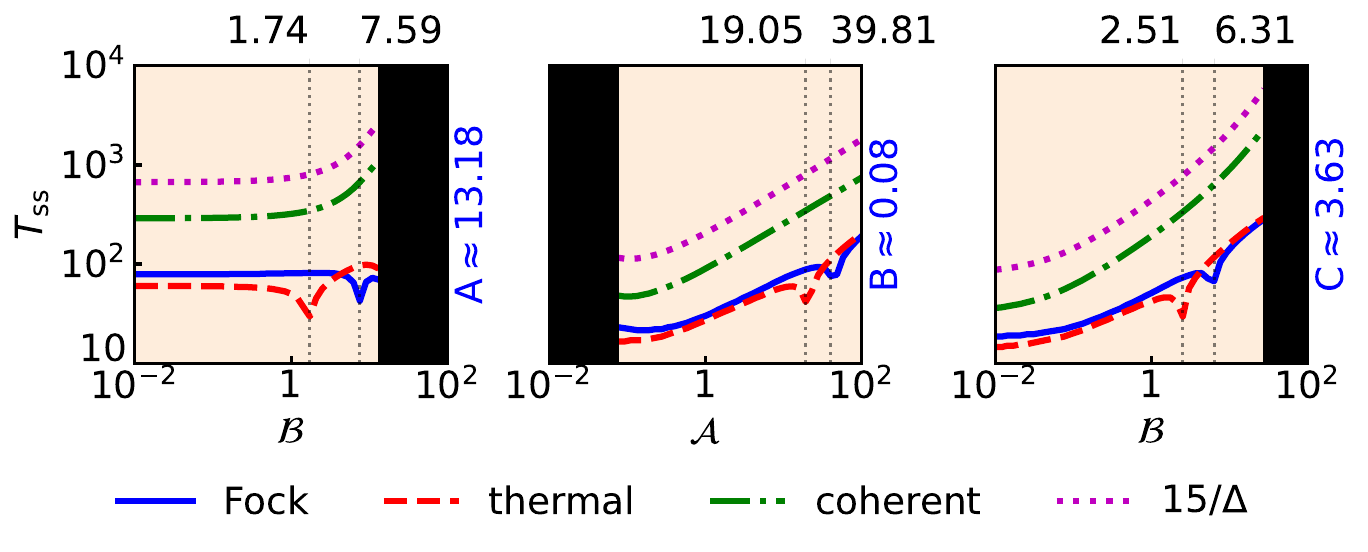}
    \caption{Slices of the second-column tile plots in Fig.~\ref{fig7} for $\woa\approx 13.18$ (40th row from the bottom), $\wob\approx 0.08$ (12th column from the left) and $\woc\approx 3.63$ (7th off-diagonal above the bifurcation boundary of $\woa=\wob$). For comparison, the corresponding values of $15/\Delta$ are presented. The values of $\Delta$ are sliced from the tile plot in Fig.~\ref{fig6} (c), which are fully valid for the slices presented here. The factor $15$ is arbitrarily chosen to shift $1/\Delta$ to the top.}
    \label{fig8}
\end{figure}

The full result is presented in Fig.~\ref{fig7}. Meanwhile, Fig.~\ref{fig8} shows selected slices of Fig.~\ref{fig7} which exhibit interesting properties. Unlike the previous results presented against the working regime points, the dependence of $\tss$ on the basis parameter is not trivial. For a given initial state, the relative values of $\tss$ between different working regime points may significantly differ depending on the exact value of $\kappa_1$. Here we merely aim to study the general features of $\tss$, so we forgo the analysis of the scaling. We use $\kappa_1=0.1$ like we do for the Liouvillian gap $\Delta$ to gain a direct comparison between $\Delta$ and $\tss$. 

For the Fock and thermal states, the behavior of $\tss$ shows a feature that is absent in $\Delta$: over some parts of the system's working regimes, $\tss$ drops to a minimum whose value is significantly smaller than the values at nearby working regime points. Additionally, when we plot $\tss$ against $\wob$ for a fixed $\woa$, we see a maximum for $\tss$ which occurs at a value of $\wob$ larger than that at which the minimum occurs. This is easier to see in thermal states for our choice of initial states, but Fock states also show this behavior. Whether or not a minimum occurs for a larger $\wob$ is beyond our knowledge. From our results, we see that this region of \emph{\revision{fast convergence regimes}} moves toward larger $\woa$ as the initial energy is increased. Furthermore, for Fock state with $\expval{\adag\aop}^{(0)}=10$, another region emerges in the small $\woa$ regime. Our findings may not capture all the properties of the steady-state regions and thus provide no insight into a pattern for their occurrence, even for a given kind of initial state. 

Nevertheless, we can logically relate the existence of the speedy regions to the system's energy dynamics, since the states exhibiting the speedy regions are diagonal. We note that the steady-state energy $\enss$ is independent of the individual values of the nonconservative parameters, as we mentioned in Section~\ref{subsection_the_SL_correspondence_regime}. A stronger energy gain than the nonlinear dissipation will shift $\enss$ toward a higher energy level, where the nonlinear dissipation is stronger. As such, despite the strong energy gain, the system struggles to reach $\enss$. Conversely, strong nonlinearity brings $\enss$ down where its effect becomes weaker. Despite the strong nonlinearity, the system also struggles to reach the low $\enss$. The \revision{fast convergence regimes} thus give the right balance for the equilibrium to be more easily attained. Additionally, where the \revision{fast convergence regimes} occur must also depend on the initial energy $\expval{\adag\aop}^{(0)}$. For example, we may consider cases where $\enss > \expval{\adag\aop}^{(0)}$ where the energy gain has to push the system's energy up against the increasingly strong nonlinear dissipation. It will take longer to reach $\enss$ the lower $\expval{\adag\aop}^{(0)}$ is. This is likewise true when $\enss < \expval{\adag\aop}^{(0)}$, in which case the progressively weakened nonlinear dissipation has to cover more energy shift opposing the energy gain to reach $\enss$. 

We find that the value of $\tss$ for a coherent state $\rho_\mathrm{coh}=\ket{\beta}\bra{\beta}$ only depends on $|\beta|$. This is not surprising: the magnitude of the coherence is the same for coherent states with the same energy, i.e.,
\begin{equation}
\left|\matrixel{m}{\rho_\mathrm{coh}}{n\neq m}\right| = e^{-\left|\beta\right|^2}\frac{\left|\beta\right|^m\left|\beta\right|^{n}}{\sqrt{m!n!}}.
\end{equation}
Additionally, in the phase space representation, we can imagine ourselves on the Wigner function peak and undergoing the dynamics in the simple-harmonic rotating frame. From this perspective, the Wigner function evolves the same way regardless of the initial angular position of the peak.

Unlike the other two states, $\tss$ for a coherent state behaves similarly to the Liouvilian gap, as clearly shown in Fig.~\ref{fig8}, which suggests that the $j=1$ term is indeed the last to vanish for a coherent state. As shown by the middle plot in Fig.~\ref{fig8}, $\tss$ also has a local minimum corresponding to the local maximum of the Liouvillian gap. However, this is not an appreciable speedup like in the cases of the other two initial states. In Section~\ref{subsection_the_classical_like_behavior}, we observe that the steady-state time for the coherence state is determined by the decay of its neighboring-level coherence. This strengthens our description of the \revision{fast convergence regimes} in terms of the energy gain: we do not see them for a coherent state since $\tss$ is determined by the coherence decay instead of the energy dynamics. 

%-------------------------------------------------------------------------------

%-------------------------------------------------------------------------------

\begin{Revision}

\section{Conclusions and Remarks}
\label{section5}

The quantum Stuart-Landau oscillator is a paradigm in the study of quantum limit cycle and synchronization. We have studied its transient dynamics and observed intriguing behaviors. We have formulated the criteria for the system to enter its classical regime, where the steady-state Wigner function is localized near its circular peak, minimizing the quantum noise. We have evaluated the equation of motion for the Wigner function and compared it to the Kramers-Moyal equation for stochastic processes. It demonstrates how the quasiprobabilistic nature of the Wigner function enables the contribution from the jumpy two-quantum dissipation to disregard the Pawula theorem, and also how this process is essential for the limit cycle to form. We have shown that this process is also responsible for the temporary increase of Wigner negativity. We have compared the evolution of a coherent state inside and outside the system's classical regime, and observed how the evolution is slow due to the slow decay of neighboring-level coherence. We have numerically evaluated the system's Lindbladian eigenvalues and observed the decrease in Liouvillian gap with stronger energy gain. Evolving the system from selected initial states, we have found that the dependence of the time to converge toward the limit cycle does not generally follow the trend of the Liouvillian gap. In particular, we have observed fast convergence regimes in the system's parameter space, over which the steady-state time is significantly reduced locally; these are not observed for the coherent states due to the lingering coherence. 
\end{Revision}

These findings provide deeper insights into the transient behavior of the quantum Stuart-Landau oscillator and a better understanding of self-sustained quantum systems in general. Nevertheless, some questions remain. We have seen that the coherence of a coherent state holds it back from reaching the steady state, leading to a steady-state time that is magnitudes larger than that of other states, such as the Fock state and the thermal state. More generally, quantum coherence slowing down the limit cycle attraction may not be desirable in some scenarios, such as when we are considering the limit cycle's robustness against coherence-generating interactions. Is there any useful way to speed up coherence decay while keeping the limit cycle unchanged? Additionally, we have seen that the \revision{fast convergence regimes} offer a significant reduction in steady-state time. Is there a general pattern for their occurrence beyond what we have discussed? Moreover, do they result in a speedy synchronization? Even if they do not, it is interesting to see whether similar regimes exist when the system is coupled with another quantum self-sustained oscillator. Additionally, we are curious whether the \revision{fast convergence regimes} can occur for nondiagonal states. This may be possible if the quantum coherence can decay faster than the time it takes for the energy to reach its steady-state value, e.g., for a cat state with many cats.

%-------------------------------------------------------------------------------
\begin{acknowledgments}
%-------------------------------------------------------------------------------
We thank Mahameru BRIN for their HPC facilities.  We acknowledge the use of the following numerical packages: \texttt{NumPy}~\cite{harris2020numpy}, \texttt{SciPy}~\cite{Scipy}, \texttt{SymPy}~\cite{sympy}, \texttt{QuTiP}~\cite{Johansson2012, Johansson2013}, \texttt{Matplotlib}~\cite{Hunter:2007matplotlib}, and \texttt{moyalstar}~\cite{moyalstarv1.1.0}.  H.M.L. is supported by a research assistantship from the BRIN Directorate for Talent Management.  The data supporting the findings of this study are available from the corresponding authors upon reasonable request.
\end{acknowledgments}

%===============================================================================
\appendix
%-------------------------------------------------------------------------------
\section{Wigner function and the system's energy}
\label{appsec_A}
%-------------------------------------------------------------------------------

As a quantum phase-space representation, the Wigner function generally represents not only quantum states but also operators. Let $\Fop$ and $\Gop$ be any two quantum operators that are not necessarily Hermitian. Then,
\begin{equation}\label{eq:wigner_overlap_formula}
    \trace{\Fop\Gop} = 2\pi\int_{-\infty}^{\infty}\int_{-\infty}^{\infty} W_{\Fop}(x,p)W_{\Gop}(x,p)\ \mathrm{d}x\ \mathrm{d}p,
\end{equation}
where $W_{\Fop}$ and $W_{\Gop}$ are the \emph{Wigner transforms} of $\Fop$ and $\Gop$, respectively obtained by replacing $\rho$ with $\Fop$ and $\Gop$ in Eq.~\eqref{eq:wigner_function}. This equation is the \emph{Wigner function overlap formula}. If one operator is the density matrix while the other is an operator, then Eq.~\eqref{eq:wigner_overlap_formula} gives the expectation value of the quantity represented by that operator. The Wigner function, like the Hilbert space vector, acts as a weight function for the expectation value. Conversely, the Wigner transform of the operator acts as a filter function that gives the Wigner function physical significance, similar to how the operator itself gives physical significance to the Hilbert state vector~\cite{leonhardt1997measuring}. We consider
\begin{subequations}
\begin{align}
    \Fop &=\rho = \sum_{\ell}\rho_{\ell \ell}\ket{\ell}\bra{\ell}+\sum_{m,n\neq m}\rho_{mn}\ket{m}\bra{n},
    \\
    \Gop &= \adag\aop = \sum_k k\ket{k}\bra{k},
\end{align}
\end{subequations}
with which the overlap formula gives $\expen$. Utilizing the superposition principle on the Wigner function, we split the Wigner transform $W_\rho (x,p)=W(x,p)$ of $\rho$ into the diagonal and off-diagonal parts:
\begin{equation}\label{eq:A3}
\begin{split}
    W(x,p) &= W_\mathrm{D}(x,p)+W_\mathrm{OD}(x,p)
    \\
    &= \sum_\ell\rho_{\ell\ell}W_{\ket{\ell}\bra{\ell}}(x,p) +\sum_{m,n\neq m}\rho_{mn}W_{\ket{m}\bra{n}}(x,p).
\end{split}
\end{equation}
We note the well-known result that the Wigner functions $W_{\ket{n}\bra{n}}$ of Fock states are rotationally invariant~\cite{leonhardt1997measuring, curtright2013concise}, which warrants the remark following Eq.~\eqref{eq:wigner_function}. As such, the lack of circular symmetry of the system's Wigner function is the consequence of a nonzero $W^\mathrm{OD}$. Next, we have
\begin{equation}\label{eq:A4}
    W_{\adag\aop}(x,p) = \sum_k k W_{\ket{k}\bra{k}}(x,p),
\end{equation}
which is rotationally invariant. We can see that the contribution from $W_\mathrm{OD}$ to $\expen$ vanishes since
\begin{equation}
\begin{split}
    0 &= 2\pi \int_{-\infty}^{\infty}\int_{-\infty}^{\infty} W_\mathrm{OD}(x,p)W_{\adag\aop}(x,p)\ \mathrm{d}x\ \mathrm{d}p 
    \\
    &= \sum_{k,m,n\neq{m}}k\rho_{mn}\trace{\ket{k}\overlap{k}{m}\bra{n}}.
\end{split}
\end{equation}
As a result,
\begin{equation}\label{eq:A6}
    \expen = 2\pi \int_{-\infty}^{\infty}\int_{-\infty}^{\infty} W_\mathrm{D}(x,p)W_{\adag\aop}(x,p)\ \mathrm{d}x\ \mathrm{d}p. 
\end{equation}
The system's energy is embodied by only the rotationally invariant part of its Wigner function corresponding to the diagonal elements---as it should be, considering that $\expen=\sum_n n\rho_{nn}$.

It is straightforward to generalize this result to any expectation value that depends only on the diagonal elements, e.g., the energy variance. Such a quantity is embodied by only the rotationally invariant part of the system's Wigner function. 

%-------------------------------------------------------------------------------
\begin{widetext}
\section{Deriving the system's Wigner function equation of motion}
\label{appsec_B}
%-------------------------------------------------------------------------------

We refer mainly to Refs.~\cite{curtright2013concise, ISAR1996} for the formalism in this section. The equation of motion for a Wigner function can be obtained by applying the Wigner transformation to the Hilbert-space equation of motion. The Wigner transformation is linear: the Wigner transform of a sum is a sum of the Wigner transforms of the summands. What about the Wigner transform of a product? The answer is the \emph{Moyal $\star$-product} of the Wigner transforms of the factors. The Moyal $\star$-product operator is defined as~\cite{curtright2013concise}
\begin{equation}
    \star = \exp\left[\frac{i}{2}\left(\overset{\leftarrow}{\partial_x}\overset{\rightarrow}{\partial_p}-\overset{\leftarrow}{\partial_p}\overset{\rightarrow}{\partial_x}\right)\right],
\end{equation}
where the over-set arrows dictate the operation direction. The $\star$-product essentially multiplies two scalars while behaving like matrix multiplication: noncommutative, associative, linear, etc. To take the Wigner transform of some function $f\left(\hat{A}_0, \hat{A}_1,\dots\right)$ of Hilbert space operators, we simply replace the operators by their Wigner transforms $W_{\hat{A}_0},W_{\hat{A}_1},\dots$ and replace all product operations with $\star$-products to obtain the ``$\star$-function'' $f_{\star}\left(W_{\hat{A}_0}, W_{\hat{A}_1},\dots\right)$. Evaluating the $\star$-product using its definition may be cumbersome. One technique is to utilize the \emph{Bopp shift}. Based on the property that
\begin{equation}
\exp\left(\hat{A}\partial_x\right)f(x) = f(x+\hat{A}),
\end{equation}
given that $\hat{A}$ is independent of $x$, we may write~\footnote{To see this more easily, consider the equivalent expression $f(x,p)\star g(x,p)=\exp\left[\frac{i}{2}\left(\partial_{x}\partial_{p'}-\partial_{p}\partial_{x'}\right)\right]f(x,p)g(x',p')|_{x'=x,p'=p}$.}
\begin{equation}
\begin{split}
    f(x,p)\star g(x,p) &= f\left(x+\frac{i}{2}\overset{\rightarrow}{\partial}_p, p-\frac{i}{2}\pright_x\right)g(x,p)
    \\
    &= f\left(x+\frac{i}{2}\pright_p,p\right)g\left(x-\frac{i}{2}\pleft_p,p\right)
    \\
    &= f\left(x,p-\frac{i}{2}\pright_x\right)g\left(x,p+\frac{i}{2}\pleft_x\right)
    \\
    &= f(x,p)g\left(x-\frac{i}{2}\pleft_p,p+\frac{i}{2}\pleft_x\right),
\end{split} 
\end{equation}
turning the $\star$-product into an ordinary product. Given this formalism, it is straightforward to obtain the Wigner transform of the Lindblad master equation:
\begin{equation}
\begin{split}
    \rho' &= -i\left[\hat{H},\rho\right] +\sum_k \gamma_k \left(\hat{O}_k\rho\hat{O}_k^\dagger-\frac{1}{2}\acomm{\hat{O}_k^\dagger\hat{O}_k}{\rho}\right)
    \\
    &\Big\downarrow \text{(Wigner transform)}
    \\
    \partial_t W &= -i\left(W_{\hat{H}}\star W - W\star W_{\hat{H}}\right) 
     +\sum_k \gamma_k\left(W_{\hat{O}_k}\star W\star W_{\hat{O}_k}^* 
    - \frac{1}{2}W_{\hat{O}_k}^*\star W_{\hat{O}_k}\star W 
     -\frac{1}{2}W\star W_{\hat{O}_k}^*\star W_{\hat{O}_k} \right),
    \\
\end{split}
\end{equation}
where we have used the property that $W_{\hat{A}^\dagger} = W_{\hat{A}}^*$. This equation can be more concisely written using the \emph{Moyal bracket} 
\begin{equation}
\moyal{f}{g} = -i\left(f\star g - g\star f\right),
\end{equation}
which is the extension of the Poisson bracket to quantum mechanics. Furthermore, as the term $\left(f\star g - g\star f\right)$ is called the ``$\star$-commutator'', we may analogously define the ``$\star$-dissipator'' $\mathcal{D}_{\star}\left(\hat{O}\right)[\rho]$ to shorthand the Wigner transforms of the Lindblad dissipators. We may thus write
\begin{equation}
    \partial_t W = \moyal{W_{\hat{H}}}{W}+\sum_k \gamma_k \mathcal{D}_{\star}\left(\hat{O}_k\right)[\rho].
\end{equation}
Our system description is given in terms of $\aop$ and $\adag$. To evaluate their $\star$-functions, we make use of the Bopp shift and the property that $W_{\hat{x}} = x$, $W_{\hat{p}} = p$ to obtain
\begin{subequations}
\begin{alignat}{2}
    W_{\aop} &= \frac{1}{\sqrt{2}}\left(W_{\hat{x}}+iW_{\hat{p}}\right) = \frac{x+ip}{\sqrt{2}}, \label{eq:B6_start}
    \\ 
    W_{\adag} &= W_{\aop}^* = \frac{x-ip}{\sqrt{2}},
    \\
    W_{\adag\aop} &= W_{\adag} \star W_{\aop} =\frac{x^2+p^2-1}{2},
   %  \\
   % W_{\aop\adag} &= W_{\aop} \star W_{\adag} = \frac{x^2+p^2+1}{2},
   %  \\
   %  W_{\aop^2} &= W_{\aop}\star W_{\aop} = \frac{x^2+2ixp-p^2}{2},
   %  \\
   %  W_{\adagn{2}\aop^2} &= W_{\adagn{2}}\star W_{\aop^2} = \frac{x^4+2p^2x^2+p^4-4x^2-4p^2+2}{4}. 
    \label{eq:B6_end}
\end{alignat}
\end{subequations}
and so on. It is then straightforward, though lengthy, to evaluate the equation of motion for the system's Wigner function using these equations. As an example, we have
\begin{equation}
\begin{split}
    W_{\adag\aop} \star W 
    % &= \frac{1}{2}\left[\left(x+\frac{i}{2}\pright_p\right)^2+\left(p-\frac{i}{2}\pright_x\right)^2-1\right]W
    % \\
    &= \left(\frac{x^2+p^2-1}{2}\right)W -\frac{i}{2}p\partial_x W +\frac{i}{2}x\partial_p W  
    \\ &\quad -\frac{1}{8}\partial_x^2W -\frac{1}{8} \partial_p^2 W,
\end{split} 
\end{equation}
and
\begin{equation}
\begin{split}
    W\star W_{\adag\aop} 
    % &= \frac{W}{2}\left[\left(x-\frac{i}{2}\pleft_p\right)^2+\left(p+\frac{i}{2}\pleft_x\right)^2-1\right]
    % \\
    &= \left(\frac{x^2+p^2-1}{2}\right)W + \frac{i}{2}p\partial_x W - \frac{i}{2}x\partial_p W 
    \\
    &\quad -\frac{1}{8}\partial_x^2 W - \frac{1}{8}\partial_p^2 W, 
\end{split}
\end{equation}
which give the simple harmonic part of the dynamics:
\begin{equation}
    \moyal{W_{\hat{H}=\adag\aop+1/2}}{W} = -p\partial_x W + x\partial_p W.
\end{equation}
Instead of evaluating the entire equation by hand, we use our home-built symbolic package \texttt{moyalstar}~\cite{moyalstarv1.1.0}, which utilizes the \texttt{SymPy} module in \texttt{Python}. After post-programming algebraic manipulation, we obtain Eq.~\eqref{eq:wigner_evo}.
\end{widetext}
\bibliography{refs}
\end{document}